\documentclass{elsarticle}
\usepackage{graphicx}
\usepackage{dcolumn}
\usepackage{bm}
\usepackage{rotating}
\usepackage{natbib}
\usepackage{txfonts}
\begin{document}
\begin{frontmatter}
\title{A study of $\Lambda$ hypernuclei within the Skyrme-Hartree-Fock 
Model }
\author[SHIMLA]{ Neelam Guleria}
\author[SHIMLA,HAMIRPUR]{ Shashi K. Dhiman}
\author[SAHA,CSSM]{Radhey Shyam}

\address[SHIMLA]{Department of Physics, H. P. University, Shimla 171005,
India} 
\address[HAMIRPUR]{University Institute of Natural Sciences and Interface
Technologies, Himachal Pradesh Technical University, Hamirpur 177001,
India}
\address[SAHA]{Saha Institute of Nuclear Physics, AF/1, Bidhan Nagar,
Kolkata 700064,India} 
\address[CSSM]{Centre for the Subatomic Structure of Matter (CSSM),
School of Chemistry and Physics, University of Adelaide, SA 5005, Australia
}

\begin{abstract}
We investigate the properties of the single $\Lambda$ hypernuclei 
within a Skyrme-Hartree-Fock (SHF) model. The parameters of the Skyrme type 
effective lambda-nucleon ($\Lambda N$) interaction are obtained by fitting 
to the experimental $\Lambda$ binding energies of hypernuclei with masses 
spanning a wide range of the periodic table. Alternative parameter
sets are also obtained by omitting nuclei below mass number 16 from the 
fitting procedure. The SHF calculations are performed for the binding 
energies of the $\Lambda$ single-particle states over a full mass range 
using the best fit parameter sets obtained in these fitting procedures and 
the results are compared with the available experimental data. The data
show some sensitivity to the parameter sets obtained with or without 
including the nuclei below mass 16. The radii of the $\Lambda$ orbits in 
the hypernuclear ground states and the $\Lambda$ effective mass in nuclear 
matter show some dependence on different parameter sets. We present 
results for the total binding energy per baryon of the hypernuclei over a 
large mass region to elucidate their stability as a function of the baryon 
number. We have also employed the our best fit $\Lambda N$ parameter sets 
to investigate the role of hyperons in some key properties of neutron stars.

\end{abstract}

\begin{keyword}
Skyrme Hatree-Fock Model \sep Structure of Hypernuclei \sep Hyperon-nucleon
effective interaction 

\PACS 21.80.+q \sep 21.10.Dr \sep 21.30.Fe
\end{keyword}

\end{frontmatter}

\section{ Introduction}

Hypernuclei provide an unique opportunity to investigate the dynamics
of the full meson and baryon SU(3) flavor octet. They are an excellent
tool to extract information on the hyperon-nucleon interaction. For a
complete understanding of the baryon-baryon interactions in terms of the
meson-exchange or quark-gluon pictures, the basic experimental data are
required also on the hyperon-nucleon interaction together with the
nucleon-nucleon interaction. Yet, relatively little is understood about
nuclei that contain one or more hyperons. It is possible that the study 
of hypernuclei may help in unraveling fundamental issues regarding
the ordinary nuclei in terms of quantum chromodynamics (QCD) which is
the theory of strongly interacting particles.

Since $\Lambda$ is the lightest among the hyperons ($m_\Lambda$ = 1.115 
GeV), the $\Lambda$ hypernuclei are the most investigated systems so 
far~\cite{dal78,pov87,yam94,gib95,kei00,gal04,has06,sai07,fin09}. Through a 
series of experimental studies involving $(K^-,\pi^-)$, $(\pi^+,K^+)$,
$(\gamma,K^+)$ and $(e,e^\prime K^+)$ reactions (see, e.g. Ref.
\cite{has06} for a recent review of the experimental scenario) the 
$\Lambda$ shell structure has been mapped out over a wide range of the 
periodic table. Systematic studies of the energy levels of light 
$\Lambda$ hypernuclei have enabled the extraction of considerable 
amount of details about the $\Lambda$N interaction. It is established 
that the spin-orbit part of this force is weaker (see, e.g. 
Ref.~\cite{gib95} and the related references cite there) than that 
of the  $NN$ system. This has also been useful in developing the 
theoretical models of the hypernuclear production reactions where the 
quantum numbers and the binding energies of $\Lambda$ single particle 
states are vital inputs~\cite{dov80,hau89,ban90,shy08,shy09,shy10}.  

For several reasons it is necessary to have information about the behavior 
of a hyperon in nuclear medium which can be provided by the heavier 
$\Lambda$ hypernuclei. The basic quantities like $\Lambda$ effective 
mass and the depth of the $\Lambda$ potential in nuclei can be obtained 
by investigating the $\Lambda$ binding energies in a wide range 
of heavier systems. These quantities are the important basic parameters 
for a realistic discussion of neutron stars~\cite{web05}. The 
information on the $\Lambda$ spin-orbit splitting in heavier 
hypernuclei is of interest because it may have contributions from the 
higher order many body effects in addition to the $\Lambda N$ two-body 
spin-dependent interaction.

It is therefore, important to develop reliable theoretical tools to
investigate the spectroscopy of the hypernuclei. From QCD point of view
hypernuclei lie in the non-perturbative low momentum regime. Therefore,
lattice QCD calculations should be the ideal tool for investigating their
structure. Indeed, first step in this direction has already been 
initiated where the scattering length and the effective range for  
$\Lambda N$ scattering have been extracted in both QCD  and  
partially-quenched QCD calculations~\cite{bea05}. However, a description 
of the detailed structure of hypernuclei is still beyond the reach of the 
lattice QCD and one has to use methods where baryons and mesons are the 
effective degrees of freedom.
                           
Both relativistic and non-relativistic descriptions have been used to 
investigate the structure of hypernuclei. The relativistic mean field 
approach has been employed in Refs.~\cite{ruf90,gle93,mar94,lom95,ver98} with 
meson-hyperon vertices adjusted in varied ways. The SU(3) symmetric field 
theories including chirality~\cite{mul99,pap99}, and the quark-meson 
coupling model~\cite{sai07,tsu98,gui08} that invokes the quark degrees 
of freedom have been developed to investigate the structure of hypernuclei. 
A density dependent relativistic hadron (DDRH) field theory was used in 
Ref.~\cite{kei00} to describe $\Lambda$ hypernuclei. In Ref.~\cite{fin09}
a in-medium chiral SU(3) dynamics has been used to study the hypernuclear
single particle spectra where a very good agreement has been achieved with 
the corresponding data. The smallness of the $\Lambda$-nuclear spin-orbit 
interaction also finds a natural explanation within this approach.
 
Among the non-relativistic approaches are the calculations based on the 
shell model picture~(see, e.g.,Ref.~\cite{mil10} for a review). Most of   
them reproduce reasonably well the measured hypernuclear states of medium 
to heavy hypernuclei using phenomenological $\Lambda$N potentials of
the Woods-Saxon type~\cite{bou76,mil88,hau89}. Several authors have also 
developed the semi-empirical mass formulas to describe the binding energy 
(BE) of the hypernuclei~\cite{iwa71,gry86,sam96} 

The Skyrme-Hartree-Fock (SHF) model, which has been a powerful tool for 
investigating the properties of nonstrange nuclei~\cite{vau72,dov72,bei75}, 
was extended to studies of the $\Lambda$ hypernuclei in Refs.
\cite{ray76,ray81}. The suitability of this approach for describing the 
$\Lambda$ hypernuclei depends heavily on the proper knowledge of the 
$\Lambda$N effective interaction, which has either been extracted from the 
microscopic methods like $G$-matrix calculations performed with J\"ulich 
and Nijmegen potentials~\cite{yam85,cug00,vid01,zho07,zho08,sch10} or by 
fitting directly to the hypernuclear data (mainly the binding energies
)~\cite{ray81,yam80,fer89}. The reliability of the latter method is related 
directly to the amount of hypernuclear data used in the fitting process. 
In previous studies, the data used for this purpose were limited to a few 
light nuclei. Therefore, there is a need to perform SHF calculations using 
the $\Lambda N$ interaction with parameters that are determined by fitting 
to a larger set of data on binding energies that is available now.  

In this paper, we study the $\Lambda$ hypernuclei within the framework of 
the Hartree-Fock method where the parameters of the Skyrme type $\Lambda$N 
effective interaction have been determined by fitting to the experimental 
binding energies of of hypernuclei with baryon numbers ranging over a  
wide mass range. In one fitting procedure, we have included nuclei with 
masses ranging between 8 to 208. In two more fittings, the nuclei with 
masses below 16 were omitted. We have also included a density dependent 
term in the $\Lambda N$ interaction in our fitting procedure as suggested 
in Ref.~\cite{lan97,mor05}. The  parameter sets having the minimum $\chi^2$ 
values in three searches have been used to calculate the binding energies 
of about 95 hypernuclear states and the results are compared with the
available experimental data. Furthermore, we have also calculated the 
total binding energy per baryon of 73 $\Lambda$ hypernuclei, and the
root mean square (RMS) radii of a few nuclei of them. Finally our
best fit parameter sets have been used to investigate some properties of
neutron stars.  

The present paper is organized as following. In section II, the 
Skyrme-Hartree-Fock method as applied to the description of the single 
$\Lambda$ hypernuclei, is briefly described and parameters involved in 
the Skyrme type $\Lambda N$ force are discussed. In section III, we 
describe the $\chi^2$ minimization method [based on the simulated 
annealing method (SAM)] that is used by us to determine the best fit 
parameters of the $\Lambda N$ interaction. In section IV, we present the 
results and discussions of the comparison of our calculations for the 
single particle energies of the $\Lambda$ states with the available 
experimental data. The results for the total binding energies per baryon 
of a larger number of single $\Lambda$ hypernuclei are also presented in
this section. We also investigate the role of our best fit $\Lambda N$ 
interactions on some properties of neutron stars. Finally, in section V, 
we present the summary and conclusions of our work. 

\section{ Formalism}

The SHF model for nonstrange nuclei has been discussed in great details 
in Refs.~\cite{vau72,dov72,bei75}. In its extention~\cite{ray76,ray81} to 
describe the $\Lambda$ hypernuclei a contribution is added to the original 
energy density functional to account for the action of the hyperon-nucleon 
force.  Thus, the total energy density functional (EDF) of a hypernucleus 
is written as a sum of two basic contributions - ${\cal E}_{N}$, which is 
the total energy density for neutron and proton \cite{vau72} and 
${\cal E}_{\Lambda}$, which is the contribution due to the presence of 
the hyperon (hyperons): 
\begin{eqnarray}
\label{edf}
{\cal E}^H_{1\Lambda}& = & {\cal E}_{N}
(\rho_n,\rho_p,\tau_n,\tau_p, {\bf J}_n, {\bf J}_p) +
{\cal E}_{\Lambda}(\rho_n,\rho_p,\rho_\Lambda,\tau_\Lambda),
\end{eqnarray}
where $\rho_q$, $\tau_q$ and $J_q$ represent the baryon, kinetic and 
spin-orbit current densities, respectively (q = $n,p ,\Lambda $). There
are some additional terms contributing to the total hypernuclear energy 
that will be discussed latter on in this section.

${\cal E}_N$ is related to the original SHF nuclear Hamiltonian density 
($H_{N}$) as 
\begin{eqnarray}
{\cal E}_{N}& =& \int d^3r H_{N}({\bf r}),
\end{eqnarray}
where H$_{N}$ is  written as~\cite{cha98,agr06}
\begin{eqnarray}
\label{H_NN}
H_{N} &=&\frac {\hbar^{2}}{2m}\tau_{N}+\frac{1}{4} t_{0}[(2+x_{0})
\rho_{N}^{2}-(2x_{0}+1)(\rho_{p}^{2}+\rho_{n}^{2})] \nonumber \\
& + & \frac{1}{24}t_{3}\rho^{\alpha}[(2+x_{3})\rho_{N}^{2}-(2x_{3}+1)
(\rho_{p}^{2}+\rho_{n}^{2})] \nonumber \\
& + & \frac{1}{8}[t_{1}(2+x_{1})+t_{2}(2+x_{2})]\tau_{N}\rho_{N}
+\frac{1}{8} [t_{2}(2x_{2}+1)-t_{1}(2x_{1}+1)](\tau_{p}\rho_{p}+
\tau_{n}\rho_{n}) \nonumber\\
& + & \frac{1}{32}[3t_{1}(2+x_1)-t_{2}(2+x_{2})](\nabla \rho_{N})^{2}-
\frac{1}{32}[3t_{1}(2x_{1}+1)+t_{2}(2x_{2}+1)][(\nabla\rho_{p})^2+
(\nabla\rho_{n})^2] \nonumber \\
& + & \frac{1}{2}W_{0}[{\bf J}_{N}.{\bf \nabla}\rho_{N}+{\bf J}_{p}.
{\bf \nabla}\rho_{p}+{\bf J}_{n}. {\bf \nabla}\rho_{n}] \nonumber \\
& - & \frac{1}{16}(t_{1}x_{1}+t_{2}x_{2}){\bf J}_{N}^{2}+\frac{1}{16}
(t_{1}-t_{2}) [{\bf J}_{p}^{2}+{\bf J}_{n}^{2}] \nonumber\\
& + & \frac{1}{2}e^{2}\rho_{p}(r) \int \frac{\rho_{p}(r')d^{3}r'}
{|\bf {r-r'}|}-
\frac{3}{4}e^2\rho_{p}(r)\Bigg(\frac{3\rho_{p}(r)}{\pi}\Bigg)^{1/3},
\end{eqnarray}
where the first term on the right hand side represents the kinetic energy.
We define, $\rho_{N} = \rho_{n}+\rho_{p},\tau_{N} = \tau_{n}+\tau_{p}$ and 
$J_{N}= J_{n}+ J_{p}$, where $n(p)$ correspond to a neutron (proton). We 
employ SLy4 Skyrme parameterization~\cite{cha98} in the calculation of the 
energy density of the hypernuclear core $^{A-1}_\Lambda Z$. This interaction 
has been widely used in studies of the nuclear structure of normal and 
neutron rich nuclei~\cite{cha03}, and the properties of nuclear matter and 
neutron stars~\cite{sto07}.

The densities $\rho_q$, $\tau_q$ and J$_q$ are expressed as 
\begin{eqnarray}
\rho_q({\bf r}) = \sum_{\beta} v_\beta \mid\phi_\beta({\bf r},q)\mid^2,\\
\tau_q({\bf r}) = \sum_{\beta} v_\beta \mid {\bf \nabla} 
                   \phi_\beta({\bf r},q)\mid^2,\\
{\bf J}_q({\bf r}) = \sum_{\beta} v_\beta \phi_{\beta}^*({\bf r},q) 
(-i{\bf \nabla} \times {\bf \sigma)} \phi_\beta({\bf r},q),
\end{eqnarray}
where $\phi_\beta({\bf r},q)$ are the wave functions of the single particle 
states, $v_\beta$ represents the occupation probability, and  $\sigma$ the 
Pauli spin matrices. The sums are taken
over all occupied states for different particles $q$. Summations over spin, 
and isospin indices are implicit.

The occupation probabilities $v_\beta$ of neutrons and protons are 
calculated by including the pairing energy functional ${\cal E}_{pair}$ in
Eq.~(1) as described in Ref.~\cite{cha03}. The BCS equations for the 
pairing probabilities are obtained by variational method with respect 
to $v_\beta$ and written as,
\begin{equation}
v_\beta^2 = \frac{1}{2}\left[ 1 - \frac{\epsilon_\beta - \mu_\beta}
{\sqrt{(\epsilon_\beta - \mu_\beta)^2 + \Delta_{q}^2}}\right],
\end{equation}
where $\epsilon_\beta$ is the single particle energy of the occupied state,
and $\mu_\beta$ is the chemical potential. The pairing gap equation has the
form $\Delta_{q} = G_q\sum_{\beta\epsilon q}\sqrt{v_\beta(1-v_\beta)}$,
where $G_q$ is as defined in Ref.~\cite{cha03}.

The energy density functional ${\cal E}_\Lambda$  is written as 
\begin{equation}
\label{eq:ELambda}
{\cal E}_{\Lambda} = \int d^3r H_{\Lambda}({\bf r}).
\end{equation}

As in Ref.~\cite{ray81}, H$_{\Lambda}$ in Eq.~\ref{eq:ELambda} has a term 
that corresponds to a Skyrme type two-body force,
\begin{eqnarray}
\label{eq:HLambdaN}
H_{\Lambda}^{\Lambda N} & = & \frac{\hbar^2}{2m_\Lambda}\tau_\Lambda 
+u_0(1+ \frac{1}{2}y_0) \rho_N \rho_\Lambda \nonumber \\
& + & \frac{1}{4}(u_1+u_2)(\tau_\Lambda \rho_N + \tau_N \rho_\Lambda)
+\frac{1}{8}(3u_1-u_2)({\bf \nabla \rho_N} \cdot {\bf \nabla \rho_\Lambda)} 
\nonumber \\
& + & \frac{1}{2}W_\Lambda({\bf \nabla \rho_N \cdot J_\Lambda} +
{\bf \nabla \rho_\Lambda \cdot \bf J_N}), 
\end{eqnarray}
In ref.~\cite{ray81} and \cite{fer89}, a second term is added to H$_{\Lambda}$
that corresponds to a zero range three-body $\Lambda NN$ force.  
\begin{eqnarray}
H_{\Lambda}^{\Lambda NN} &=& \frac{u_{3}}{4}\rho_{\Lambda}(\rho^{2}_{N}+ 
2\rho_{n}\rho_{p}).
\end{eqnarray}
This is completely analogous to the corresponding term of the nucleon 
Skyrme force proposed in Ref.~\cite{vau72}. It has been pointed out in
Refs.~\cite{gre64,bod86,tak86} that the binding energies of particularly
light hypernuclei can be better reproduced if three-body $\Lambda NN$ term is
included in the calculations. For example the overbinding problem of 
$^5_{\Lambda}$He hypernucleus has been solved by taking into account the 
$\Lambda NN$ force in Ref.~\cite{aka00}. In the Skyrme Hartree-Foch 
calculations of Ref.~\cite{fer89} it is  shown that the inclusion this term 
leads to a better reproduction of the $1p - 1d$ and $1s-1p$ levels spacings 
in several nuclei. 

In the Hartree-Fock calculations of even-even nuclei, the three-body part 
of the Skyrme interaction is equivalent to a two-body density-dependent 
interaction~\cite{vau72,bei75}. Therefore, some authors~\cite{lan97,mor05} 
have included in H$_{\Lambda}$ a term dependent on the nuclear density 
instead of the $\Lambda NN$ force,  
\begin{eqnarray}
H_{\Lambda}^{\Lambda N\rho}& = & \frac{3}{8}u_3^\prime(1+\frac{1}{2}y_3)
\rho_N^{\gamma+1} \rho_\Lambda,
\end{eqnarray}
where $\gamma$ and $y_3$ are additional parameters. It may however, be 
remarked that at least for $\gamma \ne 1$,  Eq.~(11) can not be derived 
from from any three-body force. Furthermore, the equivalence of three-body 
and the density dependent forces in the hypernuclear case is not exact 
even for $\gamma = 1$. In the limit of $\gamma =1$ and $y_3 = 0$, Eq. (11) 
does reduce to a form similar (but not strictly equal) to Eq.~(10). 
Nevertheless, using Eq.~(11) leads to a value of nuclear incompressibility 
that is closer to its experimental value~\cite{mor05}. It is stated in Ref.
\cite{lan97} that Eq.~(10) is not so adequate for representing the density
dependence of the $\Lambda N$ $G$ matrix while Eq.~(11) with a value of 
$\gamma=\frac{1}{3}$ is good enough for parameterizing the G matrix result.  
Therefore, in our calculations we have added Eq.~(11) to the Hamiltonian 
instead of the three-body force term, Eq.~(10), and have considered for 
$\gamma$ both 1 and $\frac{1}{3}$. We have used $\hbar^2/2m_\Lambda = 
17.44054$ MeV fm$^{2}$. 

Now, the single-particle wave functions $\phi_\beta ({\bf r},q)$ and 
corresponding single-particle energies $\epsilon_\beta$ are obtained by 
solving the Hartree-Fock equations with position and density dependent 
mass term
\begin{eqnarray}
\label{eq:SkHF}
\left[ -{\bf \nabla}\frac{\hbar^2}{2m_q^*({\bf r})}.{\bf \nabla} + 
V_q({\bf r}) - i W_q({\bf r})\cdot ({\bf \nabla}\times{\bf \sigma})\right]
\phi_\beta ({\bf r},q) & = & \epsilon_\beta \phi_\beta({\bf r},q),
\end{eqnarray}
where $m_q^*$ is the effective baryon mass, and $V_q$ and $W_q$ are the
central and spin-orbit terms of the mean field potential, respectively.
The central term consists of (i) a purely nuclear part ($V_{NN}$) as 
described, e.g., in Refs.~\cite{vau72,cha98}, (ii) additional field 
created by the $\Lambda$ hyperon ($V_{N}^\Lambda$) that is seen by a 
nucleon, and (iii) the whole nuclear field experienced by the 
$\Lambda$ hyperon ($V_{\Lambda}^\Lambda$). The effective mass of the 
nucleon will also have additional terms. The spin-orbit part of the 
$\Lambda N$ interaction has been ignored ($W_\Lambda$ = 0) from the start 
as is done in Ref.~\cite{ray81}. The smallness of this term has been 
supported by several microscopic calculations~\cite{bro81,pir82,tsu98}. 
The potentials $V_{\Lambda}^\Lambda$ and $V_N^\Lambda$ are written as 
\begin{eqnarray}
\label{VLN}
V_{\Lambda}^\Lambda &=& u_0(1+\frac{1}{2}y_0) \rho_N + \frac{1}{4}
(u_1+u_2)({\bf \nabla \tau_{\Lambda}\rho_{N}}+\tau_N)\nonumber\\
& + & \frac{1}{8}(3u_{1}-u_{2}) ({\bf \nabla^2 \rho_N}) -
\frac{1}{4}(3u_{1}-u_{2}) ({\bf \nabla\rho_N}/r) 
+ \frac{3}{8}u_3^\prime(1+\frac{1}{2}y_3)\rho_N^{\gamma+1},
\end{eqnarray}
and,
\begin{eqnarray}
V_N^\Lambda & = & u_0(1+\frac{1}{2}y_0) \rho_\Lambda + \frac{1}{4}
(u_1+u_2)(\tau_{\Lambda}+ {\bf \nabla \tau_N \rho_\Lambda})
 +   \frac{1}{8}(3u_1-u_2)({\bf \nabla^2 \rho_\Lambda})\nonumber \\
 &-&\frac{1}{4}(3u_{1}-u_{2}) ({\bf \nabla\rho_\Lambda}/r)
+ \frac{3}{8}u_3^\prime(1+\frac{1}{2}y_3)(\gamma + 1)
\rho_N^{\gamma}\rho_\Lambda.  
\end{eqnarray}
The spin-orbit part of the mean field has purely nucleonic contribution 
and it can be written in a straight forward standard way.

The term containing the effective mass of the single $\Lambda$ hyperon is 
expressed as,
\begin{equation}
\label{eq:effmassL}
\frac{\hbar^2}{2m_\Lambda^*} = \frac{\hbar^2}{2m_\Lambda} + \frac{1}{4}
(u_1+u_2)\rho_N.
\end{equation}
Similarly, the effective mass term for nucleon (in present of a hyperon)
is written as 
\begin{equation}
\label{eq:effmassN}
\frac{\hbar^2}{2m_{q^{\prime *}}} = \frac{\hbar^2}{2m_{q^\prime}} + 
\frac{1}{8}[t_1(2+x_1)+t_2(2+x_2)]\rho_N\nonumber \\
+\frac{1}{8} [t_2(2x_2+1)-t_1(2x_{1}+1)]\rho_{q^\prime}
+\frac{1}{4}(u_1+u_2) \rho_\Lambda, 
\end{equation}
where $q^\prime$ represents a nucleon (neutron or proton). The potentials, 
the effective masses, and the orbitals in Eq.~(\ref{eq:SkHF}) are evaluated 
alternatively until self-consistency is achieved.

The total hypernuclear energy, in a density dependent Hartree-Fock model,
includes contributions also from the $H_{NNN}$ and $H_{\Lambda N}^\rho$ 
terms.  Our total energy due to nucleon terms only includes relevant 
contributions. For the $\Lambda$ case,  we have 
\begin{eqnarray}
{\cal E}^\Lambda_R &=& \frac{3}{16}u_3^\prime \int d^3r (1+\frac{1}{2}y_3)
\rho_N^{\gamma+1}\rho_\Lambda 
\end{eqnarray}
\par
We also have to introduce to the total energy the center of mass (c.m.) 
correction arising due to the breaking of the translational invariance in 
the mean field of the Hartree-Fock theory. This is written as
\begin{eqnarray}
\label{eq:cm1}
{\cal E}_{c.m.} &=& \frac{<P_{c.m.}^2>}{2(A-n)m_N + nm_\Lambda},
\end{eqnarray}
where m$_N$ and m$_\Lambda$ are the masses of nucleon and $\Lambda$ hyperon,
respectively while n denotes the number of lambda particles (in our case 
n=1), and A is the baryon number of the hypernucleus. $P_{c.m.} = \sum_k 
\hat{p}_{k} $ is the total momentum operator in the c.m. frame, which is 
the sum of the single particle momentum operators ($\hat{p}_{k}$). Although 
the BCS state is not an eigenstate of $\hat{P}_{c.m.}$, and has vanishing 
total momentum $ <\hat{P}_{c.m.}> = 0$, it has non-vanishing expectation 
value of $<P_{c.m.}^2>$ given by,
\begin{eqnarray}
\label{eq:cm2}
< P^2_{c.m.}>& = & \sum_\alpha v^2_\alpha < \alpha\alpha\mid
{\bf p}^2\mid\alpha\alpha>\\
&-&\sum_{\alpha,\beta}v_\alpha v_{\beta}(v_\alpha v_{\beta}-u_\alpha
u_{\beta}) <\alpha\beta\mid {\bf p}_{1}.{\bf p}_{2}\mid\alpha\beta>,
\end{eqnarray}
where $u_\alpha = \sqrt{1-v_\alpha^2}$, and $\alpha$ and $\beta$ represent
the single particle states. This correction, however, is computed after
variation (i.e., {\it posteriori}). The square of the single particle
momentum operator appear only in the direct term of the c.m. correction.
The second and third terms represent the off diagonal single particle
matrix elements of the momentum operators that result from the exchange
terms in $<P_{c.m.}^2>$. The c.m. energy correction due to hyperons is
approximated by the diagonal part of the c.m. kinetic energy only.
\begin{table}
\caption{\label{table1} Experimental values of the binding energies
(in MeV) of various hypernuclei used in the $\chi^2$ minimization procedure.
 }
\vskip0.5cm
\begin{tabular}{c|c|c|c}
\hline
Hypernuclei & & BE(Expt.) & \\
 & 1s & 1p & 1d \\
\hline
$^{8}_{\Lambda}$He \cite{dav05} & $7.16\pm0.70$  & & \\
$^{9}_{\Lambda}$Li \cite{dav05} & $8.50\pm0.12$  & & \\
$^{10}_{\Lambda}$Be \cite{dav05} & $9.11\pm0.22$ & & \\
$^{10}_{\Lambda}$B \cite{has95,has96,dav05} & $8.89\pm0.12$  & & \\
$^{11}_{\Lambda}$B \cite{dav05,miy04} & $10.24\pm0.05$  & & \\
$^{12}_{\Lambda}$B \cite{dav05,ahm03} & $11.37\pm0.06$  & & \\
$^{12}_{\Lambda}$C \cite{hot01} & $10.76\pm0.19$  & & \\
$^{13}_{\Lambda}$C \cite{aji01,koh02} & $11.69\pm0.12$  & & \\
$^{16}_{\Lambda}$N \cite{cus09} & $13.76\pm0.16$  & & \\
$^{16}_{\Lambda}$O \cite{uka04,has98} & $12.50\pm0.35$  & & \\
$^{28}_{\Lambda}$Si \cite{has06,has96} & $16.60\pm0.20$ & $7.0\pm0.2$ & \\
$^{32}_{\Lambda}$S \cite{ber79} & $17.50\pm0.50$  & & \\
$^{40}_{\Lambda}$Ca \cite{tam94,chr88} & $18.70\pm1.1$  & & \\
$^{51}_{\Lambda}$V \cite{pil91,hot01} & $19.90\pm1.0$   & & \\
$^{89}_{\Lambda}$Y \cite{has06,hot01}& $23.10\pm0.50$ & $16.50\pm4.1$ & 
$9.1\pm1.3$ \\
$^{139}_{\Lambda}$La \cite{has06} & $24.50\pm1.20$ & $20.4\pm0.6$ & 
$14.3\pm0.6$ \\
$^{208}_{\Lambda}$Pb \cite{has06} & $26.30\pm0.80$ & $21.90\pm0.6$ & 
$16.8\pm0.7$ \\
\hline
\end{tabular}
\end{table}
\par
Thus, the total hypernuclear energy is given by the expression
\begin{equation}
\label{tedf}
{\cal E}^H_{1\Lambda} =  {\cal E}_{N}
(\rho_n,\rho_p,\tau_n,\tau_p, {\bf J}_n, {\bf J}_p) +
 {\cal E}_{Pair}(v_p,v_n) +
{\cal E}_{\Lambda}(\rho_n,\rho_p,\rho_\Lambda,\tau_\Lambda) -
{\cal E}^\Lambda_{R} - {\cal E}_{c.m.},
\end{equation}
where ${\cal E}_{Pair}(v_p,v_n)$ is the pairing energy density functional
as described above.

The $\Lambda$ binding energy in the SHF formalism is defined as 
\begin{equation}
B_\Lambda = {\cal E}_N^0 - {\cal E}_{1\Lambda}^H, 
\end{equation}
where ${\cal E}_{1\Lambda}^H$ and ${\cal E}_N^0$ are the total binding 
energies of the hypernucleus and the core nucleus, respectively and binding 
energy per baryon number is obtained by dividing $B_{\Lambda}$ with A 
(baryon number).

The fitting procedure to obtain the parameters of the $\Lambda$N interaction
is described in the next section.

\section{Parameterization for $\Lambda$N Skyrme Potential}

The values of parameters $u_0$, $u_1$, $u_2$, $u_3^\prime$, $y_0$ and 
$y_3$ have been determined by fitting to the experimental binding energies 
(BE) of a set of hypernuclei across the periodic table by a $\chi^2$ 
minimization procedure that is based on the SAM. This is an elegant technique
of searching for a global minimum in the hypersurface of the $\chi^2$ 
functions, which depend on the values of the parameters of the Skyrme 
interaction~\cite{agr05}. This method has been used in Ref.~\cite{agr06} to 
determine the parameters of the Skyrme type nucleon-nucleon effective 
interaction, which have been used to describe successfully the properties of 
normal and neutron rich nuclei as well as those of neutron stars.

The $\chi^2$ function is defined as
\begin{eqnarray}
\label{eq:chi2}
\chi^2 &=& \frac{1}{N_d-N_p}\sum_{i=1}^{N_d}\Bigg(\frac{M_{i}^{exp} - 
M_{i}^{th}}{\sigma_i}\Bigg)^2,
\end{eqnarray}
\par
\begin{table}
\begin{center}
\caption{\label{table2} The lower (${\bf v_{0}}$) and upper (${\bf v_{1}}$)
limits, maximum displacement ({\bf d}) and initial values (${\bf v_{in}}$)
for the Skyrme force parameters used for implementing the SAM algorithm
for minimizing the $\chi^{2}$ value that leads to the parameter set
(HP$\Lambda$2) to be shown in Table \ref{table3} .}
\begin{tabular}{|ccccc|}
\hline
Parameter & ${\bf v_{0}}$ & ${\bf v_{1}}$ & {\bf d} & ${\bf v_{in}}$\\
\hline
$u_{0}$(MeV $fm^{3}$) & -592 & -275 & 2.0 & -346 \\
$u_{1}$(MeV $fm^{5}$) & 0 & 100 & 1.0 & 60 \\
$u_{2}$(MeV $fm^{5}$) & 0 & 100 & 1.0 & 50 \\
$u_{3}^\prime$(MeV $fm^{3+3\gamma}$) & 1850 & 2300 & 3.0 & 1620 \\
$y_{0}$ & -0.7 & -0.05 & 0.01 & -0.13\\
$y_{3}$ & -1.5 & -0.01 & 0.01 & -0.38\\
\hline
\end{tabular}
\end{center}
\end{table}
\noindent
where $N_d$ is the numbers of the experimental data points and $N_p$ 
the number of fitted parameters. $\sigma_i$ stands for the theoretical 
error and $M_{i}^{exp}$ and $M_{i}^{th}$ are the experimental and the
corresponding theoretical values, respectively, of a given observable 
(which in this case is the binding energy). The values of $\chi^2$ 
depend on the Skyrme parameters being searched [in our case $u_0$, 
$u_1$, $u_2$, $u_3^\prime$, $y_0$ and $y_3$ appearing in Eqs.~(9) and (11)], 
since $M_i^{th}$ is calculated using potentials with these parameters. 
Although there are correlations in some of these parameters (e.g. $u_0$,  
$y_0$, and $u_3^\prime$, $y_3$), we have varied each one them independently -
minimization of the $\chi^2$ was the only requirement of our procedure. 
This is in complete analogy with the procedure of Refs.~\cite{agr05,agr06}.
The theoretical errors (see, Eq.~\ref{eq:chi2}) in the fitting algorithm are 
taken to be equivalent to the experimental uncertainty in the data. 
\par
The SAM is an elegant technique for optimization problems of large 
scale, in particular, where a desired global extremum is hidden among many 
local extrema. This method has been found to be an extremely useful tool for 
a wide variety of minimization problems of large non-linear systems in many 
different fields (e.g., see, Refs.~\cite{kir84,ing89,coh94}). Recently, the 
SAM was used to generate some initial trial parameter sets for the point 
coupling variant of the relativistic mean field model~\cite{bur02,bur04}.
\par
\begin{table}
\caption{\label{table3} Parameters for the $\Lambda$N Skyrme potential
derived self-consistently by fitting to the experimental binding
energies of table I for two different values of $\gamma$.}
\vskip0.5cm
\begin{tabular}{ccccccccc}
\hline
SET & $\gamma$&u$_0$& u$_1$ &u$_2$ & u$_3^\prime$ & y$_0$& y$_3$& $\chi^2$\\
& & (MeV fm$^3$)& (MeV fm$^5$) & (MeV fm$^5$) & (MeV fm$^{3+3\gamma}$)
& & & \\
\hline
HP$\Lambda$1 & 1 & -326.395 & 72.627 & -8.584 & 1746.041 & -0.223 & 
-0.389& 3.58 \\
HP$\Lambda$2 & 1 & -399.946 & 83.426 & 11.455 & 2046.818 & -0.486 & 
-0.660 & 3.06\\
HP$\Lambda$3 & 1/3 & -498.515 & 65.203 & 19.001 & 995.832 & -0.492 & 
-0.444 & 3.69 \\
HP$\Lambda$4 & 1/3 & -475.584 & 99.058 & -20.890 & 1375.172 & -0.350 & 
-0.724 & 3.76 \\
\hline
\end{tabular}
\end{table}
\par
In the SAM one needs to specify the appropriate annealing schedule together 
with the parameter space (i.e., the range of the values of the parameters) 
in which the best fit parameters are to be searched. As in Ref.
\cite{agr06}, we have employed a moderately faster Cauchy annealing 
schedule given by
\begin{eqnarray}
\label{temp}
T(k) = T_i/k,
\end{eqnarray}
where $T_i$ is the initial value of the control parameter  and T(k), 
with k= 1,2,3,....., is the control parameter at the kth step. The 
value of k is increased by unity after 120$N_p$ reconfigurations or 
12$N_p$ successful reconfigurations, whichever occurs first. The 
value of $T_i$ is taken to be 5.0, which is the same as that used 
in Ref.~\cite{agr06}. We keep on reducing the value of the control 
parameter using Eq.~(\ref{temp}) in the subsequent steps until 
$\chi^2$ becomes stationary with further reduction. Since in our case
number of parameters to be searched is rather large, we have defined
the parameter space directly in terms of the range of values each 
parameter can take. Therefore,  one of the key ingredients required for 
implementing SAM, in the present case, is to specify lower and upper 
limits for each of the parameter so that minimum is searched within these 
limits. In Table \ref{table2}, we give some details of this  procedure for 
one of the parameter sets to be shown in Table \ref{table3} that
leads to the lowest $\chi^2$. In this table, lower and upper limits of the 
values of the Skyrme parameters are denoted by $\bf{v_o}$ and $\bf{v_1}$ 
in each case, respectively. In the third column, $\bf{d}$ represents the 
maximum displacement allowed in a single step for a given Skyrme parameter 
during the reconfiguration process (see Refs.~\cite{agr06} and \cite{agr05} 
for more details).  $\bf{v_{in}}$, in the last column, shows the initial 
values of the Skyrme parameters used as a starting point for the SAM.
\par
\begin{table}
\caption{\label{Ex-bepa}Experimental binding energy (in MeV) of various 
states of known single lambda hypernuclei with masses $\geq$ 16 used in the 
alternative fitting procedure.}
\vskip0.5cm
\begin{tabular}{c|c|c|c|c|c|c}
\hline
Hypernuclei & && BE(Expt.) && \\
 & 1s & 1p & 1d & 1f & 1g \\
\hline
$^{16}_{\Lambda}$N  & $13.76\pm0.16$  & $2.84\pm0.16$&&& \\
$^{16}_{\Lambda}$O  & $12.50\pm0.35$  & $1.85\pm0.05$&&& \\
$^{28}_{\Lambda}$Si & $16.60\pm0.20$ & $7.0\pm0.2$ &&& \\
$^{32}_{\Lambda}$S  & $17.50\pm0.50$  & & &&\\
$^{40}_{\Lambda}$Ca & $18.70\pm1.1$  & &&& \\
$^{51}_{\Lambda}$V  & $19.97\pm0.13$   & $11.28\pm0.6$ &&& \\
$^{89}_{\Lambda}$Y  & $23.10\pm0.50$ & $16.50\pm4.1$ & $9.1\pm1.3$ &
$2.3\pm1.2$ & \\
$^{139}_{\Lambda}$La& $24.50\pm1.20$ & $20.4\pm0.6$ & $14.3\pm0.6$
& $8.0\pm0.6$ & $1.5\pm0.6$ \\
$^{208}_{\Lambda}$Pb& $26.30\pm0.80$ & $21.90\pm0.6$ & $16.8\pm0.7
$ & $11.7\pm0.6$ & $6.6\pm0.6$ \\
\hline
\end{tabular}
\end{table}
\par
We first applied this procedure to fit the BE data listed in Table 
\ref{table1}. In our search procedure, we used two values for the parameter
$\gamma$, 1 and  $\frac{1}{3}$.  For each $\gamma$, 2 parameter sets, 
generated by following the fitting procedure discussed above, are shown 
in Table 3. The $\chi^2$ values for each set is given in the last 
column of this table. Sets HP$\Lambda$1 and HP$\Lambda$2 have $\gamma = 1$ 
while  HP$\Lambda$3 and HP$\Lambda$4 have $\gamma = \frac{1}{3}$. We note 
that parameter set HP$\Lambda$2 has the lowest $\chi^2$ value among of all 
the sets. With $\gamma = \frac{1}{3}$, set HP$\Lambda$3 provides the lowest
$\chi^2$. We have used the set HP$\Lambda$2 in further calculations  
of the hypernuclear properties unless specified otherwise. In neutron star 
studies, however, the set HP$\Lambda3$ has also been used. The 
different sets of parameters shown in Table \ref{table3}, lead to a depth 
of the $\Lambda$N mean field of about 28.00$\pm$0.58 MeV 
at the nuclear matter density ($\rho_0$) of 0.16$\pm0.01$fm$^{-3}$.
 
It could, however, be argued that hypernuclei with baryon number below 16 
are too light to be used in the Skyrme Hartree-Fock fitting procedure, even 
through the Hartree-Fock method itself has been used to calculate the 
properties of light nuclei like He and Be by several authors (see, e.g. 
Ref.~\cite{gib69} and \cite{zho07}). We have, therefore, also performed our 
fitting procedure by excluding all the nuclei with masses less than 16. 
Furthermore, since, experimental binding energies of $1f$ orbit in 
$^{89}_{\Lambda}$Y and $1f$ and $1g$ orbits in $^{139}_{\Lambda}$La and 
$^{208}_{\Lambda}$Pb are also available (see, e.g., Refs.~\cite{hot01,has06}),
we have included them also in our this alternative fitting. The BEs of the 
$1s$ and $1p$ levels of the hypernucleus $^{16}_{\Lambda}$N have recently 
become available from the study of the $(e,e^\prime K)$ reaction in the 
Jefferson Laboratory~\cite{cus09} while those for the $^{16}_{\Lambda}$O 
hypernucleus are given in Ref.~\cite{pil91}. These binding energies together 
with those of the other heavier hypernuclear systems are shown in 
table~\ref{Ex-bepa}.

We have performed two sets of fits - one by including BE data of the 
$^{16}_{\Lambda}$N hypernucleus together with those the rest of the nuclei 
shown in table~\ref{Ex-bepa} but excluding the $^{16}$O data, and the another 
where the BE data of $^{16}$O system are included but those of $^{16}$N are 
excluded. The resulting parameters having the minimum $\chi^2$ values, are 
shown in table~\ref{alt-para} with $gamma$ parameter of 1 and $1/3$. The 
parameter sets N$\Lambda$1 and N$\Lambda$2 include the binding energies
of $^{16}_{\Lambda}$N while  O$\Lambda$1 and O$\Lambda$2 include those of 
$^{16}_{\Lambda}$O. We note from table~\ref{alt-para} that the minimum 
$\chi^2$ values obtained in the alternative fitting procedure are even 
lower than than those shown in table~\ref{table3}, with the set O$\Lambda$1
having the lowest $\chi^2$ value among all the sets.  We have used the 
parameter sets HP$\Lambda$2 (that corresponds to minimum $\chi^2$ in 
table~\ref{table3}), N$\Lambda$1 and O$\Lambda$1 of table \ref{alt-para} 
in our calculations on the hypernuclear properties presented in
the next section.
 
\begin{table}
\caption{\label{alt-para} Parameters for the $\Lambda$N Skyrme potential
derived self-consistently by fitting to the experimental binding
energies of table~\ref{Ex-bepa} for two different values of $\gamma$. 
The sets N$\Lambda$1 and N$\Lambda$2 are obtained by including the BEs of 
$^{16}_{\Lambda}$N and those of the rest of the nuclei but excluding the 
BE of $^{16}$O while O$\Lambda$1 and O$\Lambda$2 corresponds to the fits 
achieved by including BEs of $^{16}_{\Lambda}$O and the rest but excluding the
BE of $^{16}$N}.
\vskip0.5cm
\begin{tabular}{ccccccccc}
\hline
SET & $\gamma$&u$_0$& u$_1$ &u$_2$ & u$_3^\prime$& y$_0$& y$_3$& $\chi^2$\\
& & (MeV fm$^3$)& (MeV fm$^5$) & (MeV fm$^5$) & (MeV fm$^{3+3\gamma}$)
& & & \\
\hline
N$\Lambda$1 & 1 & -253.3250 &147.1264 &-83.5843 & 1684.9876 & 0.5802 & 0.4831 & 2.30 \\
O$\Lambda$1 & 1 & -236.5835 &116.8704 &-112.8812&1453.3493 & 0.1271 & -0.3110 & 1.82 \\
N$\Lambda$2 & 1/3 & -518.620 & 82.0944 &-19.9772&1190.1894&-0.1392 & 0.3126 & 2.51 \\
O$\Lambda$2 & 1/3 & -417.7593 &1.5460 &-3.2617&1102.2221 &-0.3854 & -0.5645 & 1.92 \\
\hline
\end{tabular}
\end{table}

\section{Results and Discussions}
\subsection{$\Lambda$ single particle fields}
\begin{figure}[t]
\includegraphics[scale=0.65]{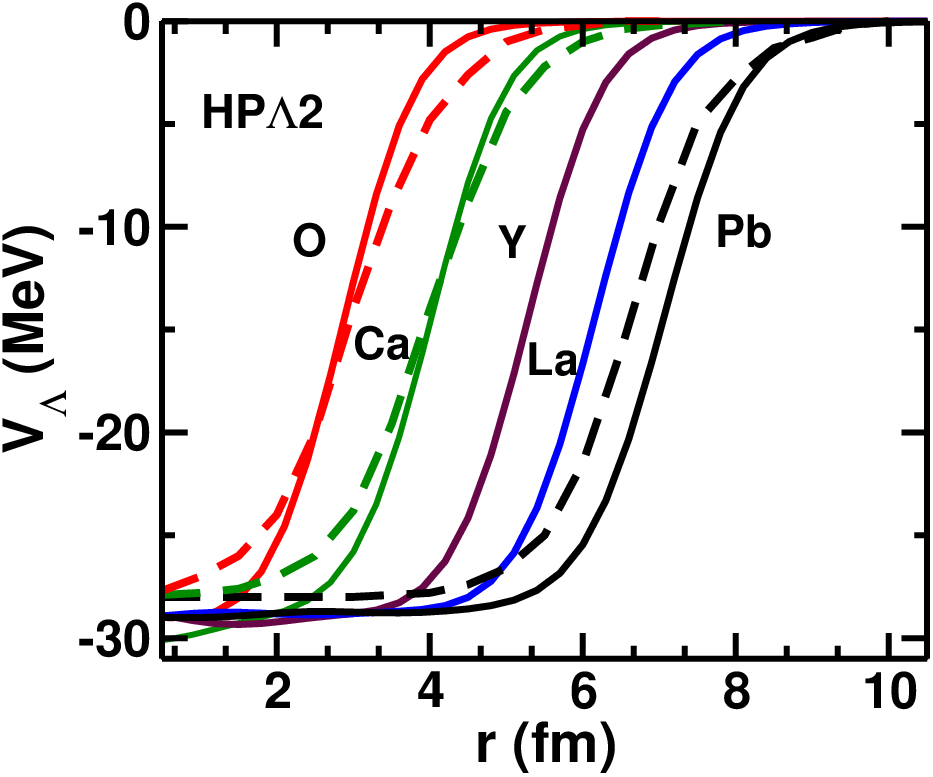}
\caption{[color online] The solid lines show the coordinate space 
SHF mean field potential $V_{\Lambda}$ in hypernuclei $^{16}_\Lambda$O, 
$^{40}_\Lambda$Ca, $^{89}_\Lambda$Y, $^{139}_\Lambda$La, and 
$^{208}_\Lambda$Pb as marked near each line. The corresponding fields 
generated in an empirical Woods-Saxon parameterization (that are taken 
taken from Ref.~\protect\cite{mil88}) are also shown for O, Ca and Pb 
nuclei by dashed lines near the lines showing the corresponding Hartree-Fock
$V_\Lambda$.} 
\label{fig1} 
\end{figure}
\par
The existence of the ground states of $\Lambda$ hypernuclei is related 
to the single particle field ($V_\Lambda$) for the $\Lambda$ hyperon.  
In the SHF model the $\Lambda N$ force contributes to the field $V_\Lambda$ 
in proportional to various powers of $\rho_N$. It would be instructive 
to compare this quantity calculated within the Hartree-Fock theory 
($V_\Lambda^{HF}$) with those obtained with a phenomenological 
parameterization of ($V_\Lambda^{ph}$) e.g. of Ref.~\cite{mil88} for 
hypernuclei with masses varying in a broad range. In Fig.~\ref{fig1}, we 
show $V_{\Lambda}^{HF}$ [calculated using Eq.~(\ref{VLN}) with the 
parameter set HP$\Lambda$2], for hypernuclei $^{16}_{\Lambda}$O,  
$^{40}_{\Lambda}$Ca, $^{89}_{\Lambda}$Y, $^{139}_{\Lambda}$La and 
$^{208}_{\Lambda}$Pb. We also show in this figure the $V_\Lambda^{ph}$ 
obtained within the parameterization of Ref.~\cite{mil88} for O, Ca and 
Pb systems - the corresponding potentials are actually taken from this
reference.  We notice that the central depths of potentials 
$V_{\Lambda}^{HF}$ are around 29 MeV irrespective of the nuclear mass. In 
comparison, $V_\Lambda^{ph}$ are somewhat less attractive. This result is 
in contrast to that shown in Ref.~\cite{cug00} where the $V_{\Lambda}^{HF}$ 
was found to be shallower than $V_\Lambda^{ph}$. Use of a different 
$\Lambda N$ force could be one reason for this difference.  

For the $Pb$ nucleus, the $V_{\Lambda}^{HF}$ extends to the larger values 
of $r$ in comparison to $V_\Lambda^{ph}$. Similar trend was seen in 
Ref.~\cite{cug00}. This result has been attributed to the presence of the 
effective mass $m^*$ in the HF equation which mocks up for some finite 
range effects in such approaches. This results is also in agreement with
the observations made in Ref.~\cite{mil88} where it was shown that the 
presence of the $\rho^2$ terms in the $\Lambda$ potentials [Eqs.~(13) and 
(14)] leads to an interaction with an increased half value radius as 
compared to that of the simple Woods-Saxon potential for heavier nuclei. 
However, in lighter systems the $\rho^2$ dependent term of the 
potential seems to reduce the effective radius of the potential.   
\par
\subsection{Binding energies of $\Lambda$ single-particle states}
\begin{figure}[t]
\includegraphics[scale=0.65]{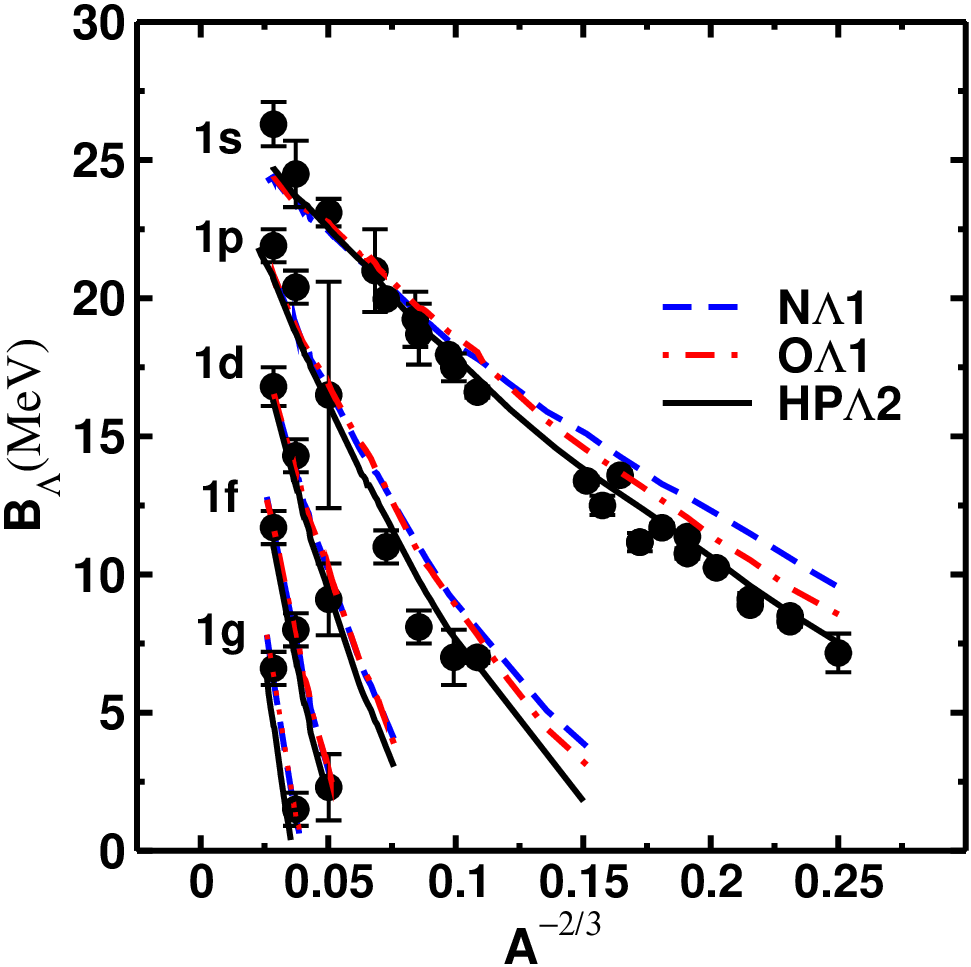}
\label{fig2}
\caption{[color online] The comparison of experimental and calculated 
binding energies of $1s$, $1p$, $1d$, $1f$, and $1g$ $\Lambda$ states as 
a function of $A^{-2/3}$ where A is the baryon number of the hypernucleus. 
In each case the solid, dashed and dashed-dotted lines correspond to the 
results obtained with $\Lambda N$ interaction parameter sets of 
HP$\Lambda$2, N$\Lambda$1 and O$\Lambda$1, respectively.
} 
\end{figure}
\par
In Fig.~2, we show a comparison of the calculated and the experimental 
$\Lambda$ binding energies ($B_\Lambda$) of $1s$, $1p$, $1d$, $1f$ and $1g$ 
shells of a number of hypernuclei with masses ranging from 8 to 208. In 
these calculations, we have used SLy4 for the $NN$ interaction and 
the parameter sets HP$\Lambda$2, N$\Lambda$1 and O$\Lambda$1 for the 
$\Lambda N$ interaction. We note that there is an overall agreement between
our calculations performed with HP$\Lambda$2 set and the experimental data 
in both the low mass and the high mass regions. However, sets O$\Lambda$1
and N$\Lambda$1 tend to overbind the $1s$ and also to some extent the $1p$ 
$\Lambda$ orbitals for lighter nuclei. This effect is relatively more
stark with the set N$\Lambda$1. However, since the set HP$\Lambda$2 has 
been obtained by including the BE of the light nuclei as well in the fitting
procedure, it is not surprising that it provides a better description of
the $B_\Lambda$ of the $1s$ and $1p$ orbitals of the lighter systems. 

For heavier nuclei all the three sets describe the $B_\Lambda$ equally well.
In a marked contrast to the SHF results of Ref.~\cite{cug00}, we observe the 
underbinding only for the $1s$ single particle orbital of $^{208}$Pb 
hypernucleus. Moreover, even for this case the magnitude of the 
underbinding is much smaller in our calculation as compared that of 
Ref.~\cite{cug00}. In the SHF calculations presented in Ref.~\cite{lan97}, 
underestimations of varying magnitudes were observed for the single 
particle energies of both $s$ and $p$ orbitals of the heavier systems with 
most of the $\Lambda N$ interactions used.  In the density dependent 
hadron field theory calculations of Ref.~\cite{kei00}, the single particle 
energies were underpredicted by factors of up to 2.5 for hypernuclei with 
masses below $^{28}_\Lambda$Si. This was attributed to the enhancement of 
surface effects in the lighter nuclei that are not accounted for properly 
in their calculations.
\par
Looking at the fact that the parameter sets HP$\Lambda$2  provides a good  
overall description of the single particle spectra of the $\Lambda$ 
hypernuclei which in some ways improves upon previous SHF calculations 
performed with different parameterizations, the predictions of our model 
can be used with relatively more confidence in calculations of the 
hypernuclear production cross sections via e.g.  $(\pi^+,K^+)$~\cite{shy10},
or $(\gamma,K^+)$~\cite{shy08,shy09} reactions for those cases where 
experimental information may not be available about the binding energies 
of the $\Lambda$ states.  

In table {\ref{table6}}, we show the effective $\Lambda$ hyperon masses 
and the root mean square (rms) radii ($r_\Lambda$) of the ground state 
$\Lambda$ orbits of various hypernuclei calculated at the nuclear saturation 
density ($n_{0}$) with various $\Lambda N$ parameterizations obtained 
by us. This provides additional constraints on different parameter sets. 
First of all we note that  $m^{*}_{\Lambda}$/$m_{\Lambda}$ obtained with 
interactions having $\gamma$ value of (1/3) (last 4 rows of this table) are 
slightly different from those involving $\gamma = 1$. The effective mass is 
largest with the parameter sets O$\Lambda$1 and O$\Lambda$2. Furthermore, 
these sets give the lowest $r_\Lambda$ for various hypernuclei. It is also 
of interest to note that $r_\Lambda$ calculated with our parameter sets are 
systematically larger than those obtained with Skyrme $\Lambda N$ 
interactions of Ref.~\cite{lan97}. It should however be mentioned that some 
uncertainty in the rms radii originates from the choice of the NN potential 
which was also different in Ref.~\cite{lan97}.
\begin{table}
\caption{\label{table6}The $\Lambda$ effective mass (calculated at saturation
density $\rho_{0}$) and rms radii of ground state $\Lambda$ orbitals of
various hypernuclei.}
\vskip 1cm
\begin{tabular}{c|c|c|c|c|c|c|c|c|c}
\hline
SET & $m^{*}_{\Lambda}/m_{\Lambda}$ & $r_{\Lambda}$($^{9}_\Lambda$Li)
& $r_{\Lambda}$($^{12}_\Lambda$C) & $r_{\Lambda}$($^{16}_\Lambda$O)
& $r_{\Lambda}$($^{28}_\Lambda$Si) & $r_{\Lambda}$($^{40}_\Lambda$Ca)
& $r_{\Lambda}$($^{51}_\Lambda$V) & $r_{\Lambda}$($^{139}_\Lambda$La)
& $r_{\Lambda}$($^{208}_\Lambda$Pb) \\
 & & (fm)& (fm) & (fm)& (fm)& (fm)&(fm)&(fm)&(fm) \\
\hline
HP$\Lambda$1 & 0.88 & 3.16 & 2.96 & 2.98 & 3.09 & 3.27 & 3.42 & 4.10 & 4.75 \\
HP$\Lambda$2 & 0.85 & 3.17 & 2.98 & 3.00 & 3.12 & 3.31 & 3.46 & 4.17 & 4.81 \\
N$\Lambda$1  & 0.88 & 2.94 & 2.88 & 2.92 & 3.11 & 3.31 & 3.46 & 4.14 & 4.81 \\
O$\Lambda$1  & 0.99 & 2.94 & 2.78 & 2.82 & 2.93 & 3.11 & 3.27 & 3.87 & 4.54 \\
HP$\Lambda$3 & 0.86 & 3.19 & 2.98 & 2.99 & 3.09 & 3.27 & 3.43 & 4.11 & 4.75 \\
HP$\Lambda$4 & 0.87 & 3.19 & 2.98 & 3.00 & 3.10 & 3.28 & 3.43 & 4.12 & 4.75 \\
N$\Lambda$2  & 0.88 & 2.95 & 2.87 & 2.92 & 3.09 & 3.28 & 3.44 & 4.09 & 4.76 \\
O$\Lambda$2  & 1.00 & 2.93 & 2.74 & 2.80 & 2.90 & 3.07 & 3.23 & 3.82 & 4.49 \\
\hline
\end{tabular}
\end{table}
\par
\subsection{Total binding energy per baryon of hypernuclei}

Although the  hyperon contribution to the total hypernuclear energy is 
relatively small, a systematic study of the total binding energies per baryon 
($BE/A$) of hypernuclei could be useful because it sheds light on the pattern 
of the stability of hypernuclei across the periodic table. Looking at the 
experimental data one notices that the BE/A  of the medium mass hypernuclei 
is larger than those of the lighter and the heavier ones. This indicates that 
hypernuclei in this region may be more stable. In the SHF framework, the 
total binding energy of a hypernucleus of baryon number A can be obtained 
from Eq.~(22).
\begin{table}
\caption{\label{bepa1}Binding energy per baryon number (BE/A) of various
known single lambda hypernuclei and corresponding values calculated by RMF
method~\cite{lan09} are also listed for comparison.}
\begin{tabular}{cccccc}
\hline
Hypernuclei & BE/A & BE/A (RMF) & Hypernuclei & BE/A & BE/A (RMF) \\
 & (MeV) & (MeV) & & (MeV) & (MeV) \\
\hline
$^{9}_{\Lambda}$Be & 7.115 & - & $^{62}_{\Lambda}$Ni & 8.773 & 8.839 \\
$^{9}_{\Lambda}$B & 5.906 & - & $^{63}_{\Lambda}$Ni & 8.773 & 8.844 \\
$^{14}_{\Lambda}$C & 8.631 & - & $^{64}_{\Lambda}$Ni & 8.768 & 8.853 \\
$^{14}_{\Lambda}$N & 7.076 & - & $^{65}_{\Lambda}$Ni & 8.764 & 8.865 \\
$^{15}_{\Lambda}$N & 8.523 & - & $^{86}_{\Lambda}$Kr & 8.727 & 8.778 \\
$^{20}_{\Lambda}$Ne & 7.386 & 7.635 & $^{87}_{\Lambda}$Kr & 8.695 & 8.796 \\
$^{21}_{\Lambda}$Ne & 7.579 & 7.714 & $^{87}_{\Lambda}$Rb & 8.745 & 7.680 \\
$^{24}_{\Lambda}$Mg & 7.624 & 7.723 & $^{88}_{\Lambda}$Rb & 8.719 & 7.587 \\
$^{27}_{\Lambda}$Al & 8.184 & - & $^{88}_{\Lambda}$Sr & 8.754 & 7.657 \\
$^{33}_{\Lambda}$S & 8.666 & - & $^{89}_{\Lambda}$Sr & 8.734 & 7.564 \\
$^{36}_{\Lambda}$S & 8.579 & 8.637 & $^{90}_{\Lambda}$Y & 8.739 & 7.510 \\
$^{37}_{\Lambda}$S & 8.579 & 8.704 & $^{90}_{\Lambda}$Zr & 8.740 & 7.545 \\
$^{38}_{\Lambda}$Ar & 8.585 & 8.597 & $^{91}_{\Lambda}$Zr & 8.731 & 7.455 \\
$^{39}_{\Lambda}$Ar & 8.634 & 8.728 & $^{92}_{\Lambda}$Mo & 8.675 & 7.378 \\
$^{48}_{\Lambda}$Ca & 8.863 & 8.855 & $^{93}_{\Lambda}$Mo & 8.678 & 7.291 \\
$^{49}_{\Lambda}$Ca & 8.749 & 8.880 & $^{112}_{\Lambda}$Sn & 8.471 & 6.925 \\
$^{50}_{\Lambda}$Ti & 8.823 & 8.843 & $^{113}_{\Lambda}$Sn & 8.479 & 6.961 \\
$^{51}_{\Lambda}$Ti & 8.747 & 8.904 & $^{114}_{\Lambda}$Sn & 8.487 & 6.996 \\
$^{52}_{\Lambda}$V & 8.751 & 8.909 & $^{115}_{\Lambda}$Sn & 8.495 & 7.032 \\
$^{54}_{\Lambda}$Fe & 8.725 & 8.774 & $^{116}_{\Lambda}$Sn & 8.502 & 7.030 \\
$^{55}_{\Lambda}$Fe & 8.715 & 8.897 & $^{117}_{\Lambda}$Sn & 8.509 & 7.028 \\
$^{58}_{\Lambda}$Ni & 8.666 & 8.856 & $^{118}_{\Lambda}$Sn & 8.516 & 7.027 \\
$^{59}_{\Lambda}$Ni & 8.705 & 8.847 & $^{119}_{\Lambda}$Sn & 8.523 & 7.026 \\
$^{60}_{\Lambda}$Ni & 8.738 & 8.841 & $^{120}_{\Lambda}$Sn & 8.530 & 7.026 \\
$^{61}_{\Lambda}$Ni & 8.763 & 8.838 & $^{121}_{\Lambda}$Sn & 8.538 & 7.026 \\
\hline
\end{tabular}
\end{table}
\par
In Tables (\ref{bepa1}) and (\ref{bepa2}) we present the results of our 
SHF calculations (done with the HP$\Lambda$2 force) for $BE/A$ for 
73 hypernuclei with masses in the range of 9 to 211. The results obtained 
with sets N$\Lambda$1 and O$\Lambda$1 are almost the same. We note that for 
the lighter systems ($ A <$ 28), the stability of a hypernucleus depends on 
that of the nucleus where a neutron is replaced by the hyperon. For example, 
$^{9}_{\Lambda}$Be is more stable than $^{9}_{\Lambda}$B, and $^{14}_\Lambda$C 
is more stable than $^{14}_\Lambda$N. For hypernuclei with baryon number 
in the range of 30-95 the $BE/A$ is around 8.7 MeV and it decreases gradually 
with further increase in the mass number. For comparison purpose the
results of a relativistic mean field (RMF) model taken from Ref.
\cite{lan09} are also presented wherever available. RMF results for the 
$BE/A$ are given only in this reference. We note that that for nuclei 
with baryon number in excess of 87 the RMF model underbinds the hypernuclei. 
However, the difference seen between SHF and RMF energies could possibly 
result from the differences in the nuclear energies predicted by the two 
models. Therefore, one should be careful in interpreting the comparison of 
the results produced by these two mean field models. 
\begin{table}
\caption{\label{bepa2}Same as table \ref{bepa1}.}
\vskip0.5cm
\begin{tabular}{cccccc}
\hline
Hypernuclei & BE/A & BE/A (RMF) & Hypernuclei & BE/A & BE/A (RMF) \\
 & (MeV) & (MeV) & & (MeV) & (MeV) \\
\hline
$^{122}_{\Lambda}$Sn & 8.545 & 7.026 & $^{141}_{\Lambda}$Ce & 8.204 & 6.874 \\
$^{123}_{\Lambda}$Sn & 8.551 & 7.027 & $^{141}_{\Lambda}$Pr & 8.337 & 6.823 \\
$^{124}_{\Lambda}$Sn & 8.554 & 7.028 & $^{142}_{\Lambda}$Pr & 8.339 & 6.855 \\
$^{125}_{\Lambda}$Sn & 8.552 & 7.030 & $^{142}_{\Lambda}$Nd & 8.321 & 6.800 \\
$^{132}_{\Lambda}$Sn & 8.415 & 7.066 & $^{143}_{\Lambda}$Nd & 8.325 & 6.834 \\
$^{133}_{\Lambda}$Sn & 8.391 & 7.074 & $^{144}_{\Lambda}$Sm & 8.280 & 6.753 \\
$^{136}_{\Lambda}$Xe & 8.392 & 6.949 & $^{145}_{\Lambda}$Sm & 8.289 & 6.790 \\
$^{137}_{\Lambda}$Xe & 8.381 & 6.968 & $^{209}_{\Lambda}$Pb & 7.886 & 6.726 \\
$^{138}_{\Lambda}$Ba & 8.375 & 6.886 & $^{210}_{\Lambda}$Bi & 7.854 & 6.711 \\
$^{139}_{\Lambda}$Ba & 8.369 & 6.911 & $^{210}_{\Lambda}$Po & 7.826 & 6.689 \\
$^{140}_{\Lambda}$La & 8.362 & 6.893 & $^{211}_{\Lambda}$Po & 7.833 & 6.696 \\
$^{140}_{\Lambda}$Ce & 8.352 & 6.845 & & & \\

\hline
\end{tabular}
\end{table}

\subsection{Properties of a neutron star}

In order to calculate neutron-star properties, it is necessary to have 
an Equation of State (EOS) linking pressure to the total energy density of 
the dense matter. At densities closer to the nuclear saturation 
density ($n_0 = 0.16 fm^{-3}$), the matter is mostly composed of 
neutrons, protons and leptons (electrons and muons) in $\beta$ equilibrium.
As density increases, new hadronic degrees of freedom may appear. Hyperons
are one of them as the equilibrium conditions in neutron stars make the 
formation of hyperons energetically favorable. The role of hyperons on the 
neutron star properties has been studied by several authors (see, e.g.
Refs.~\cite{gle85,gle91,bal97,bal00,vid00,sch06,sch11,wei12,whi12}).
\begin{figure}[t]
\includegraphics[scale=0.65]{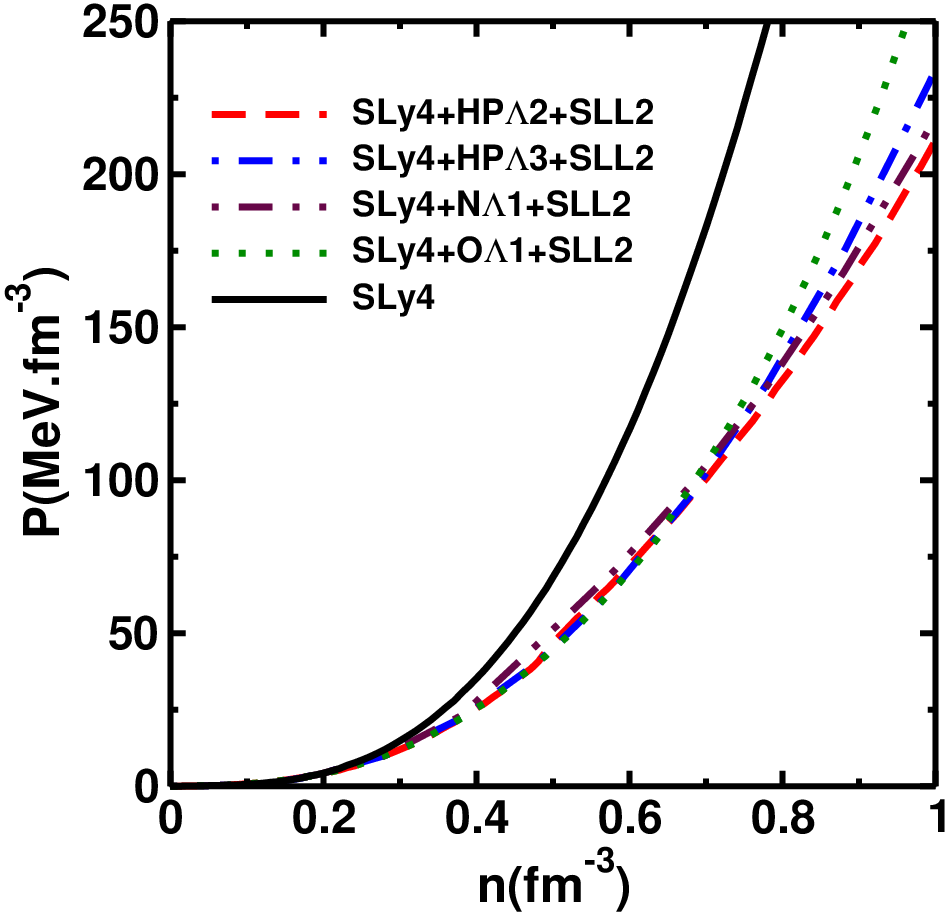}
\label{fig3}
\caption{[color online] The equation of state (pressure as a function of
baryon number density) obtained by using $\Lambda N$ interactions 
HP$\Lambda$2 (dashed line), HP$\Lambda$3 (dashed-dotted line), N$\Lambda$1 
(dashed-double dotted line) and O$\Lambda$1 (dotted line).} 
\end{figure}
\par
In this section, we employ the best fit $\Lambda N$ interactions obtained 
by us to discuss the implications of hyperons on the EOS and the structure of 
neutron stars. Unlike several other authors, we have included only the
$\Lambda$ hyperons into our calculations as this study is restricted to 
testing the interactions obtained in this work. In that sense our work 
may appear to be less complete in comparison to many previous studies where 
include other hyperons were included as well. Nevertheless, in all likelihood
the $\Sigma N$ interaction is repulsive because no stable $\Sigma$ 
hypernucleus other than that of mass 4, is known to exist. Therefore, 
$\Sigma$ appears at much higher densities~\cite{wei12} as compared to 
$\Lambda$. The $\Xi$ hyperon, on the other hand, could appear at densities 
comparable to those of $\Lambda$. However, there is considerable amount of 
uncertainty about the strength of $\Xi N$ interaction as no bound 
$\Xi$ hypernucleus has been detected so far. The threshold of the appearance
of the $\Xi$ hyperons is pushed to higher densities with increasing 
$\Xi N$ potential.  

In our calculations of the EOS, we have closely followed the methods
reported in Refs.~\cite{cha97,rik03,mor05,agr06}. To our energy density 
functional [Eq.~(1)] we have also added a contribution corresponding to 
the $\Lambda \Lambda$ interaction as at higher densities the results are 
sensitive to this term, for which the energy density functional is taken
from Ref.~\cite{mor05} with parameters corresponding to their set SLL2.
The energy density includes the rest energies of the matter constituents. 
The leptonic contribution to the energy density is calculated as discussed 
in Refs.~\cite{mor05,agr06} (see also~\cite{bay71}). For this sector 
the energy densities corresponding to electrons and muons are written as
\begin{eqnarray}
\label{lepe}
{\cal E}_l & = & \frac{1}{\pi^2}\int_0^{k_f^l} k^2 \sqrt{k^2+m_l^2} dk,
\end{eqnarray} 
where $l = e$ (electron) or $\mu$ (muon) and $k_f^l$ is the corresponding
Fermi momentum. For a given baryon density the values of the Fermi 
momenta for neutrons, protons, $\Lambda$, electron and muon can be 
obtained by requiring that neutron star matter is in $\beta$ equilibrium.
This leads to the following equations for the equality of the chemical 
potentials (represented by $\mu$ in the following)
\begin{eqnarray}
\label{chep}
\mu_n -\mu_p & = & \mu_e,\,\,\,\,\,\, \mu_\mu = \mu_e \nonumber \\
\mu_n + m_n & = & \mu_\Lambda + m_\Lambda,
\end{eqnarray}
where chemical potentials are defined as
\begin{eqnarray}
\mu_j & = & \frac{\partial{\cal E}}{\partial{n_j}}
\end{eqnarray}
where ${\cal E}$ is total energy density and $n_j$ the particle number
density. 
\begin{figure}[t]
\includegraphics[scale=0.65]{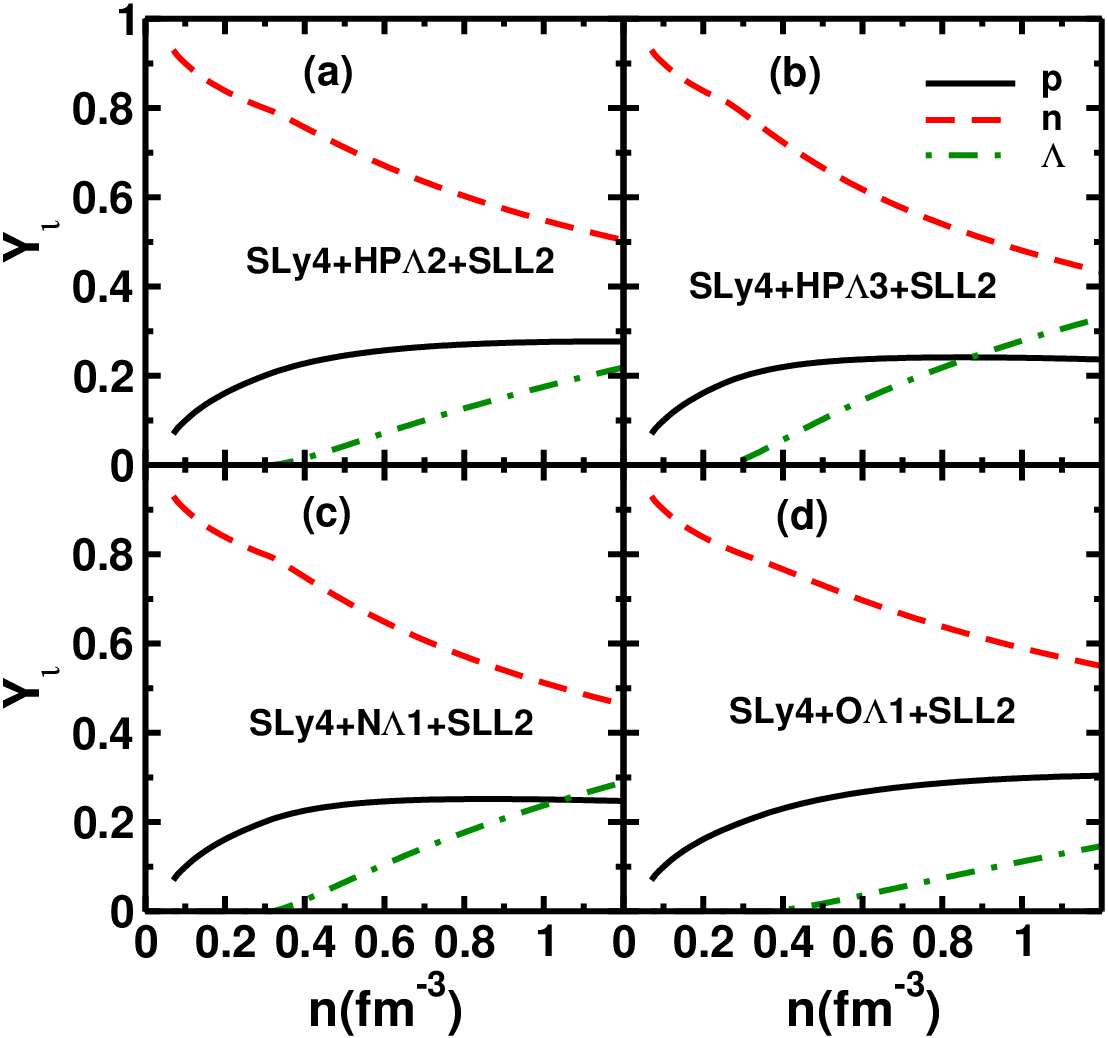}
\label{fig4}
\caption{[color online] Particle fractions obtained with using $\Lambda N$ 
interactions HP$\Lambda$2 (a), HP$\Lambda$3 (b), N$\Lambda$1 (c) and 
O$\Lambda$1 (d).}
\end{figure}
\par
The total baryon number density is $n_b = n_n + n_p$ and the charge 
neutrality requires $n_p = n_e + n_\mu$, where $n_e$ and $n_\mu$ are the 
number densities of electrons and muons, respectively. These equations 
combined with Eq.~\ref{chep} give the particle fraction 
$Y_j = \frac{n_j}{n_b}$. The EOS is defined by the expressions 
\begin{eqnarray}
\rho(n_b) = \frac{{\cal E}(n_b)}{c^2}, \,\,\,\,\,\, P(n_b) = n_b^2 
\frac{d({\cal E}/n_b)}{dn_b},
\end{eqnarray}
where $\rho(n_b)$ is the mass density of the matter. 

In Fig.~3 we show the equation of state calculated with $\Lambda N$ 
interactions HP$\Lambda$2, HP$\Lambda3$, N$\Lambda$1, and O$\Lambda$1. 
It is to be noted that in each case the inclusion of hyperons makes the 
EOS much softer with respect to that of the pure nucleonic case. Since 
hyperons can be accommodated in the lower momentum states, their kinetic 
energies are decreased which leads to the softening of the EOS. We see 
that the degree of softness of the EOS obtained with HP$\Lambda$2 and 
N$\Lambda$1 interactions are almost identical. However, with HP$\Lambda$3 
the softness is comparatively smaller and with O$\Lambda$1 the softening 
of the EOS is relatively the lowest.   
\begin{figure}[t]
\includegraphics[scale=0.65]{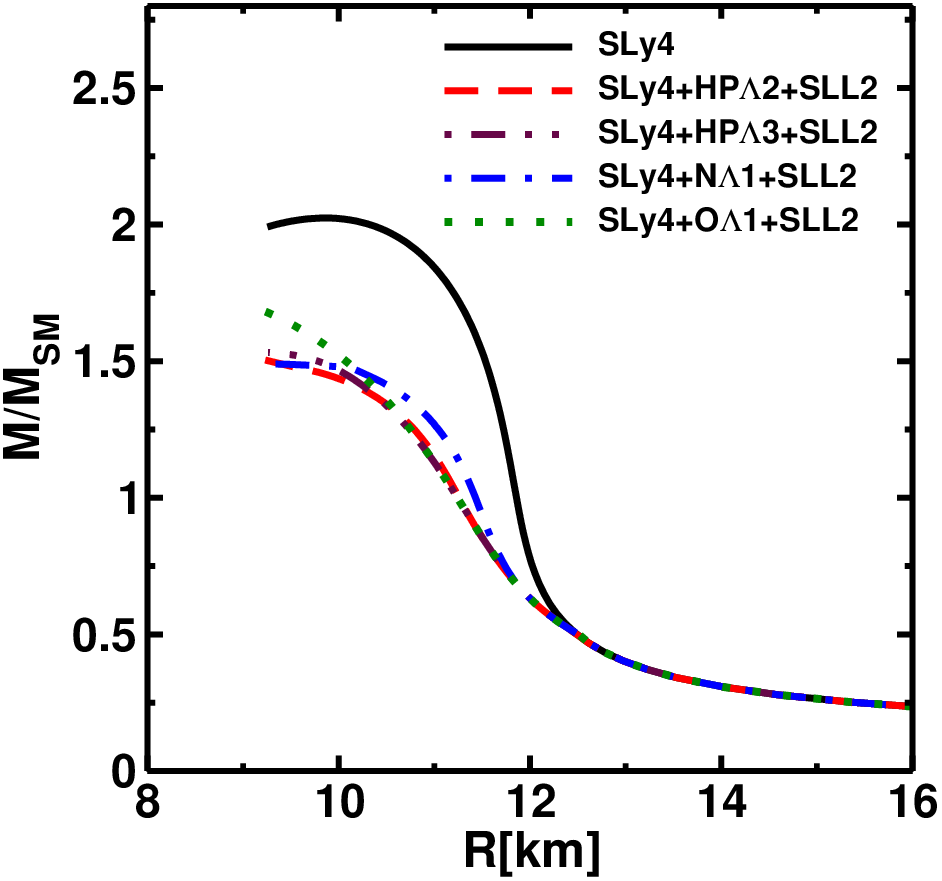}
\label{fig5}
\caption{[color online] Mass radius relation for neutron stars obtained 
with EOS shown in Fig.~3 with  interactions HP$\Lambda$2, HP$\Lambda$3, 
N$\Lambda$1 and O$\Lambda$1.}
\end{figure}
\par
In Fig.~4, we show the particle fraction $Y_i$ as a function of the 
total baryonic density. It is clear from this figure that neutrons are 
by far the most dominant object of the matter. One further notices that 
while in panels (a), (b) and (c) $\Lambda$ hyperon appears at about 
twice the nuclear saturation density, in panel (d) they start only after 
about 2.5$n_0$. They are also less numerous in this case 
as compared to those in other 3 panels. This is the consequence of the fact 
that with the interaction O$\Lambda$1 the softness introduced to the nucleon
only EOS is the least. The rise of hyperon fraction, after their formation, 
is quickest with HP$\Lambda$3 and N$\Lambda$1 interactions as compared to 
that seen with with other interactions.
\par
Further consequences of the different nature of the EOS softening with
the introduction of the $\Lambda$ hyperon with different $\Lambda N$ 
interactions can be seen on the their effect on the neutron star masses. 
To obtain the relation between neutron star mass and its radius, we have 
solved the  Tolmann-Oppenheimer-Volkoff equation 
\begin{eqnarray}
\label{tov1}
\frac{dP}{dr} & = & -\frac{G{\cal E}(r)M(r)}{r^2c^2}\bigl[1 + \frac{P(r)}
{{\cal E}(r)}\bigr]\bigl[1 + \frac{4\pi r^3 P(r)}{M(r)c^2}\bigr]\nonumber \\
 & \times & \bigl[1 - \frac{2GM(r)}{rc^2}\bigr]^{-1},
\end{eqnarray}
and 
\begin{eqnarray}
\label{tov2}
\frac{dM(r)}{dr} & = & \frac{4\pi r^2 {\cal E}(r)}{c^2},
\end{eqnarray} 
where $P$, ${\cal E}$, and $M$ are the pressure, energy density and 
gravitational mass of the neutron star, respectively. These quantities
depend on the distance $r$ from the center. Eqs.~\ref{tov1} and \ref{tov2}
can be solved from the knowledge of the initial value of the pressure $P$ 
at the center ($r$ =0) and using $M(0) =0$ and the relationship of 
$P$ and ${\cal E}$ (EOS). 

The predicted masses of the neutron star [measured in solar masses $M_\odot$]
as a function of radius (in kilometer[km]) are shown in Fig.~5 using EOS 
of Fig.~3. We note that with EOS obtained with interactions HP$\Lambda$2, 
HP$\Lambda3$, and N$\Lambda$1 maximum neutron star mass is similar (around 
1.5 $M_\odot$). In Fig.~5, $M_\odot$ is depicted as M$_{SM}$.
However, with the EOS corresponding to the interaction O$\Lambda$1 the 
maximum mass is about 1.75 M$_\odot$. This result can be understood from 
the fact that a stiffer EOS leads to a larger neutron star maximum mass. 
It may be mentioned here that if $\Lambda \Lambda$ interaction is switched 
off the maximum neutron star mass fails to reach to even its "canonical" 
value of 1.4 $M_\odot$.

\section{Summary and conclusions}

In summary, we have used an extended Skyrme Hartree-Fock model to describe 
the properties of the $\Lambda$ hypernuclei. New parameterizations for the 
Skyrme type $\Lambda N$ force have been obtained by fitting to the 
experimental binding energies of about 20 $\Lambda$ hypernuclear orbitals 
with the baryon number ranging from 8 to 208 (the best fit parameter set is 
termed at HP$\Lambda$2). We have also performed the fittings by excluding 
the binding energies of nuclei with masses below 16. In this case two sets 
of parameters were obtained:  in the first one (termed as N$\Lambda$1) 
the binding energies of the $^{16}_{\Lambda}$O hypernucleus were excluded 
but those of the $^{16}_{\Lambda}$N hypernucleus measure recently in a 
$(e,e^\prime K^+)$ experiment were include in the fitting procedure, while 
in the second one (termed as O$\Lambda$1) the reverse was done. The 
fitting method uses an elegant $\chi^2$ minimization method based on the 
simulated annealing method. 

These three sets of the best fit parameters were used to calculate the 
binding energies of $1s$, $1p$, $1d$, $1f$ and $1g$ shells of a number of 
hypernuclei. We find that calculations performed with the set 
HP$\Lambda$2 provide the best agreement with the experimental binding 
energies in the entire range of mass values and for all the orbitals. 
The set O$\Lambda$1 produces binding energies, which overestimate somewhat 
the corresponding experimental data of the $1s$ and to a lesser extent of
the $1p$ orbitals for the lighter nuclei. This overestimation is  
larger in case of the set N$\Lambda$1. However, all the three sets produce 
equally good agreement with the experimental data for heavier nuclei  
for all the orbitals.
  
Except for one case [$1s$ orbital of $^{208}$Pb nucleus], our calculations 
do not underbind the heavier systems. This is a marked improvement over the 
similar previous Hartree-Fock studies of $\Lambda$ hypernuclei where binding 
energies of orbitals of several heavier nuclei were underpredicted. 
Furthermore, the root mean square radii of heavier systems are predicted to 
be larger than those obtained in previous SHF calculations done with 
different $\Lambda N$ forces.

We made a systematic study of the mass dependence of the total binding energy 
per baryon of 73 $\Lambda$ hypernuclei spanning the entire range of the 
periodic table. It is observed that hypernuclei with masses in the range 
of 30-95 are more stable than those lying in other regions. For lighter 
systems some of the hypernuclei are more stable than their immediate neighbors 
on both sides.  

We have also tested our best fit $\Lambda N$ interactions to investigate
the role of hyperons in the neutron star sector. In these studies, we used 
the same $NN$ effective interaction ($SLy4$) together with HP$\Lambda$2, 
HP$\Lambda$3, N$\Lambda$1 and O$\Lambda$1 effective $\Lambda N$ interactions.
Inclusion of the $\Lambda \Lambda$ interaction was also found to be 
necessary in these studies - we took the corresponding effective Hamiltonian 
from the work of Lanskoy~\cite{lan98}. It is noted that inclusion of 
the $\Lambda$ hyperon makes the neutron star equation of state softer. 
However, while the sets HP$\Lambda$2, HP$\Lambda$3 and N$\Lambda$1 lead 
the softness of the similar magnitude, that obtained with the set 
O$\Lambda$1 is lesser. Furthermore, with this set the $\Lambda$ hyperon 
appear at a relatively higher density and are relatively less numerous 
as compared to other three sets. 

The maximum neutron star mass obtained with sets HP$\Lambda$2, HP$\Lambda$3
and N$\Lambda$1 is about 1.5$M_\odot$, while that with the set O$\Lambda$1 
is 1.75$M_\odot$. This is the direct consequence of the relatively stiffer
EOS in this case. We remark that these results obtained with our best 
fit $\Lambda N$ effective interactions reproduce quantitatively all the 
features that have been observed in several different models such as 
non-relativistic Brueckner-Hartree-Fock and relativistic mean field 
calculations. Nevertheless, an explanation of some of the recent neutron 
star observations~\cite{dem10} such as pulsar PSR J1614-2230 with a mass of 
1.97$\pm$0.04$M_\odot$ is still beyond the scope of the present Skyrme 
type of model unless the EOS is even stiffer than what has been achieved 
with the interaction O$\Lambda$1.   
    
One of the authors (NG) would like to thank the theory division of the  
Saha Institute of Nuclear Physics, Kolkata for financial support and 
hospitality during a visit. She is also grateful to Himachal Pradesh 
University, Shimla for providing her a Ph.D. scholarship. Useful
discussions with Prof. A.~W.~Thomas are gratefully acknowledged. This work 
has been partly supported by the University of Adelaide and the Australian
Research Council through grant FL0992247(AWT).


\begin {thebibliography}{99}

\bibitem{dal78}
R.~H.~Dalitz and A.~Gal, Ann. Phys. {\bf 116} (1978) 167.

\bibitem{pov87}
B.~Povh, Progr. Part. Nucl. Phys., {\bf 18} (1987) 183.

\bibitem{yam94}
Y.~Yamamoto, T.~Matoba, H.~Himeno, K.~Ikeda, and S.~Nagata,
Prog. Theor. Phys. Suppl. {\bf 117} (1994) 361.

\bibitem{gib95}
B.~F.~Gibson and E.~V.~Hungerford~III, Phys. Rep. {\bf 257} (1995) 349.

\bibitem{kei00}
C.~M.~Keil, F.~Hoffmann, and H.~Lenske, Phys. Rev. C {\bf 61} (2000)
064309.

\bibitem{gal04}
A.~Gal, Prog. Theor. Phys. Suppl. {\bf 156} (2004) 1.

\bibitem{has06}
O.~Hashimoto and H.~Tamura, Prog. Part. Nucl. Phys. {\bf 57} (2006) 564.

\bibitem{sai07}
K.~Saito, K.~Tsushima, and A.~Thomas, Prog. Part. Nucl. Phys. {\bf 58} 
(2007) 1.

\bibitem{fin09}
P.~Finelli, N.~Kaiser, D.~Vretenar and W.~Weise, Nucl. Phys. {\bf A 831}
(2009) 163.

\bibitem{dov80}
C.~B.~Dover, L.~Ludeking, G.~E.~Walker, Phys. Rev. C {\bf 22} (1980) 2073.

\bibitem{hau89}
R.~Hausmann and W.~Weise, Nucl. Phys. {\bf A 491} (1989) 598

\bibitem{ban90}
H.~Bando, T.~Motoba and J. Zofka, Int. J. Mod. Phys. A {\bf 5} (1990) 4021.

\bibitem{shy08}
R.~Shyam, H.~Lenske and U.~Mosel, Phys. Rev. C {\bf 77} (2008) 052201.

\bibitem{shy09}
R.~Shyam, K.~Tsushima and A.~W.~Thomas, Phys. Lett. {\bf B 676} (2009) 51.

\bibitem{shy10}
S.~Bender, R.~Shyam and H.~Lenske, Nucl. Phys. {\bf A 839} (2010) 51.

\bibitem{web05}
F.~Weber, Prog. Part. Nucl. Phys., {\bf 54} (2005) 193;
J.~Schaffner-Bielich, J. Phys. G {\bf 31} (2005) S651.

\bibitem{bea05}
S.~R.~Beane, P.~F.~Bedaque, A.~Parreno, and M.~J.~Savage, Nucl. Phys. A 
{\bf 747} (2005) 55.

\bibitem{ruf90}
M.~Rufa, J.~Schaffner, J.~Maruhn, H.~Stocker, W.~Greiner, P.~G.~Reinhard,
Phys. Rev. C {\bf 42} (1990) 2469.

\bibitem{gle93}
N.~K.~Glendenning, D.~Von-Eiff, M.~Haft, H.~Lenske, and M.~K.~Weigel, 
Phys. Rev. C, {\bf 48} (1993) 889.

\bibitem{mar94}
J. Mares and B.K. Jennings, Phys. Rev. C {\bf 49} (1994) 2472.

\bibitem{lom95}
R.~J.~Lombard, S.~Marcos and J. Mares, Phys. Rev. C {\bf 51} (1995) 1784.

\bibitem{ver98}
D.~Vretenar, W.~Poschl, and G.~A.~Lalazissis, and P.~Ring, 
Phys. Rev. C, {\bf 57} (1998) 1060.  

\bibitem{mul99}
H.~M\"uller, Phys. Rev. C {\bf 59} (1999) 1405.

\bibitem{pap99}
P.~Papazoglou, S.~Schramm, J.~Schaffner-Bielich, H.~St\"ocker,
W.~Greiner, Phys. Rev. C {\bf 57} (1998) 2576.

\bibitem{tsu98}
K.~Tsushima, K.~Saito, J.~Heidenbauer, A.~W.~Thomas, Nucl. Phys. 
{\bf A630} (1998) 691.

\bibitem{gui08}
P.~M.~Guichon, A.~W.~Thomas, and K.~Tsushima, Nucl. Phys. {\bf A 814}
(2008) 66.

\bibitem{mil10}
D.~J.~Millener, Nucl. Phys. {\bf A 835} (2010) 11c and references therein.

\bibitem{bou76}
A.~Bouyssy and J.~H\"ufner, Phys. Lett. {\bf B 64} (1976) 276;
A.~Bouyssy, Phys. Lett. {\bf B 84} (1979) 41.

\bibitem{mil88}
D.~J.~Millener, C.~B.~Dover, and A.~Gal, Phys. Rev. C {\bf 38} (1988) 
2700.
\bibitem{iwa71}
S.~Iwao, Prog. Theo. Phys., {\bf 46} (1971) 1407.

\bibitem{gry86}
M.~Grypeos, G.~Lalazissis and S.~Massen, Nucl. Phys. {\bf A 450} (1986)
283c.

\bibitem{sam96}
C.~Samanta, P.~Roy~Chowdhury and D.~N.~Basu, J. Phys. G:Nucl. Part. Phys.
{\bf 32} (2006) 363.

\bibitem{vau72}
D.~Vautherin and D.~M.~Brink, Phys. Lett. {\bf 32B} (1970) 149;
D.~Vautherin and D.~M.~Brink, Phys. Rev. C {\bf 5} (1972) 626;
D.~Vautherin, Phys. Rev. C {\bf 7} (1973) 296.

\bibitem{dov72}
C.~B.~Dover and Nguyen~Van~Giai, Nucl. Phys. {\bf A 190} (1972) 373;
 
\bibitem{bei75}
M.~Beiner, H.~Flocard, Nguyen~Van~Giai, and P.~Quentin, Nucl. Phys. 
{\bf A 238} (1975) 29.

\bibitem{ray76}
M.~Rayet, Ann. of Phys. {\bf 102} (1976) 226.

\bibitem{ray81}
M.~Rayet, Nucl. Phys. {\bf A 367} (1981) 381.

\bibitem{yam85}
Y.~Yamamoto and H. Bando, Prog. Theo. Phys. Suppl. {\bf 81} (1985) 42;
Y. Yamamoto, T. Motoba, H. Himeno, K. Ikeda and S. Nagata, Prog. Theo.
Phys. Suppl. {\bf 117} (1994) 361.

\bibitem{cug00}
J.~Cugnon, A.~Lejeune, and H.~J.~Schulze, Phys. Rev. C {\bf 62} (2000) 
064308.

\bibitem{vid01}
I.~Vida\~na, A.~Polls, A.~Ramos, and H.-J.~Schulze, Phys. Rev. C {\bf 64}
(2001) 044301.

\bibitem{zho07}
Xian-Rong Zhou, J.-J.~Schulze, H. Sagawa, Chen-Xu Wu, En-Guang Zhao, 
Phys. Rev. C {\bf 76} (2007) 034312.

\bibitem{zho08}
Xian-Rong Zhou, A.~Polls, H.-J.~Schulze, and I.~Vida\~na, Phys. Rev. C
{\bf 78} (2008) 054306.

\bibitem{sch10}
H.-J.~Schulze, Nucl. Phys. {\bf A 835} (2010) 19.

\bibitem{yam80}
Y.~Yamamoto, H.~Bando and J.~Zofka, Prog. Theor. Phys. {\bf 80} (1988) 757.

\bibitem{fer89}
F.~Fernandez, T.~Lopez-Arias and C.~Prieto, Z. Phys. A {\bf 334} (1989)
349.

\bibitem{lan97}
D.~E.~Lanskoy and Y.~Yamamoto, Phys. Rev. C {\bf 55} (1997) 2330.

\bibitem{mor05}
L.~Mornas, Eur. Phys. J. A {\bf 24} (2005) 293.

\bibitem{cha98}
E.~Chabanat, P.~Bonche, P.~Haensel, J.~Meyer, and R.~Schaeffer, Nucl. Phys. 
A {\bf 635} (1998) 231.

\bibitem{agr06}
B.~K.~Agrawal, S.~K.~Dhiman, and R.~Kumar, Phys. Rev. C {\bf 73}  
(2006) 034319. 
 
\bibitem{cha03}
S.~S.~Chandel, S.~K.~Dhiman, and R.~Shyam, Phys. Rev. C {\bf 68} 
(2003) 054320.

\bibitem{sto07}
J.~R.~Stone and P.~G.~Reinhard, Prog. Nucl. Part. Phys. {\bf 58} (2007) 587.

\bibitem{gre64}
H.~S.~Green, Nucl. Phys. {\bf 57} (1964) 483.

\bibitem{bod86}
A.~R.~Bodmer and Q.~N.~Usmani, Nucl. Phys. {\bf A 450} (1986) 257c

\bibitem{tak86}
S.~Takeuchi and K. Shimizu, Phys. Lett. {\bf B 179} (1986) 197.

\bibitem{aka00}
Y. Akaishi et al., Phys. Rev. Lett. {\bf 84} (2000) 3539.

\bibitem{bro81}
R.~Brockmann and W.~Weise, Nucl. Phys. {\bf A 355} (1981) 365.

\bibitem{pir82}
H.~J.~Pirner, B.~Povh, Phys. Lett. {\bf B 114} (1982) 308.

\bibitem{agr05}
B.~K.~Agrawal, S.~Shlomo, and V.~K.~Au, Phys. Rev. C {\bf 72} (2005) 014310.
 
\bibitem{dav05}
D.~H.~Davis, Nucl. Phys. {\bf A 754} (2005) 3c.

\bibitem{has95}
T.~Hasegawa and et al., Phys. Rev. Lett. {\bf 74} (1995) 224.

\bibitem{has96}
T.~Hasegawa and et al., Phys. Rev. C {\bf 53} (1996) 1210.

\bibitem{miy04}
Y.~Miyura et al., Acta Phys. Polon., {\bf B35} (2004) 1019.

\bibitem{ahm03}
M.~Ahmed et al., Phys. Rev. C {\bf 68} (2003) 064004.

\bibitem{hot01}
H.~Hotchi and et al., Phys. Rev. C {\bf 64} (2001) 044302.

\bibitem{aji01}
S.~Ajimura et al., Phys. Rev. Lett. 86 (2001) 4255.

\bibitem{koh02}
H.~Kohri et al., Phys. Rev. C {\bf 65} (2002) 034607.

\bibitem{cus09}
F.~Cusanno and et al., Phys. Rev. Lett. {\bf 103} (2009) 202501.

\bibitem{uka04}
M.~Ukai and et al., Phys. Rev. Lett. {\bf 93} (2004) 232501.

\bibitem{has98}
O.~Hashimoto et al., Nucl. Phys. {\bf A 639} (1998) 93c.

\bibitem{ber79}
R.~Bertini and et al., Phys. Lett. B, {\bf B 83} (1979) 306.

\bibitem{tam94}
H.~Tamura et al., Prog. Theo. Phys. {\bf 117} (1994) 1.

\bibitem{chr88}
R.~E.~Chrien and et al., Nucl. Phys. {\bf A 478} (1988) 705.

\bibitem{pil91}
P.~H.~Pile and et al., Phys. Rev. Lett. {\bf 66} (1991) 2585.

\bibitem{kir84}
S.~Kirkpatrik, J. Stat. Phys. {\bf 34} (1984) 975.

\bibitem{ing89}
L.~Ingber, Math. Comput. Modeling {\bf 12} (1989) 967.

\bibitem{coh94}
B.~Cohen, Master's Thesis, Tel-Aviv University (1994) (unpublished).

\bibitem{bur02}
T.~Burvenich, D.~G.Madland, J.~A.~Maruhn and P.~-G.~Reinhard, 
Phys. Rev. C {\bf 65} (2002) 044308.

\bibitem{bur04}
T.~Burvenich , D.~G.~Madland, and P.~G.~Reinhard, 
Nucl. Phys. {\bf A 744} (2004) 92.

\bibitem{lan09}
Lan~Mi-Xiang, Li~Lei, Ning~Ping-Zhi, Chinn. Phys. Lett. {\bf 26} (2009)
072101.

\bibitem{gle85}
N.~K.~Glendenning, Astrophys. J. {\bf 293} (1985) 470.

\bibitem{gle91}
N.~K.~Gledenning and S. Moszkowski, Phys. Rev. Lett. 67 (1991) 2414.

\bibitem{bal97}
S.~Balberg and A.~Gal, Nucl. Phys. {\bf A 625} (1997) 435.

\bibitem{bal00}
M.~Baldo, G.~F.~Burgio and H.-J.~Schulze, Phys. Rev. C {\bf 61} (2000)
055801.

\bibitem{vid00}
I.~Vidana, A.~Polls, A.~Ramos, L.~Engvik, and M.~Hjorth-Jensen,
Phys. Rev. C {\bf 62} (2000) 035801.

\bibitem{sch06}
H.-J.~Schulze, A. Polls, A. Ramos and I. Vidana, Phys. Rev. C {\bf 73}
(2006) 058801.

\bibitem{sch11}
H.-J.~Schulze, and T.~Rijken, Phys. Rev. C {\bf 84} (2011) 035801.

\bibitem{wei12}
S.~Weissenborn, D.~Chatterjee and J.~Schaffner-Bielich, Nucl. Phys.
{\bf A 881} (2012) 62. 
 
\bibitem{whi12}
D.~L.~Whittenbury, J.~D.~Carroll, A.~W.~Thomas, K.~Tsushima, and 
J.~R.~Stone, arXiv:1204.2614 [nucl-th].

\bibitem{cha97}
E.~Chabanat, P.~Bonche, P.~Haensel, J.~Meyer and R.~Schaeffer, Nucl. Phys.
{\bf A 627} (1997) 710.

\bibitem{rik03}
J.~Rikovska Stone, J.~C.~Miller, R.~Koncewicz, P.~D.~Stevenson, and 
M.~R.~Strayer, Phys. Rev. C {\bf 68} (1993) 034324.

\bibitem{bay71}
G.~Baym, C.~Pethick and P. Sutherland, Astrophys. J. {\bf 170} (1971) 299.

\bibitem{dem10}
P.~B.~Demorest, T.~Pennucci, S.~M.~Ransom, M.~S.~E.~Roberts and 
J.~W.~T.~Hessels, Nature {\bf 467} (2010) 1081.

\bibitem{gib69}
B.~F.~Gibson, A. Goldberg, and M. S. Weiss, Phys. Rev. {\bf 181} (1969) 1486.

\bibitem{lan98}
D.~E.~Lanskoy, Phys. Rev. C {\bf 58} (1998) 3351.

\end{thebibliography}
\end{document}